\documentclass{sig-alternate-2013}

\usepackage{graphicx}
\usepackage{balance}  
\usepackage{xcolor}
\usepackage{algorithm}
\usepackage{algpseudocode}
\usepackage{enumitem}
\usepackage{times}
\usepackage{amsmath}
\usepackage{xspace}
\usepackage{booktabs}
\usepackage{mathrsfs}
\usepackage{wrapfig}

\usepackage{microtype}
\usepackage[nocompress]{cite}

\newcommand{\disable}[1]{#1}

\disable{
\usepackage{subcaption}
\captionsetup[figure]{font=bf}
\captionsetup[subfigure]{font=normalfont}
}

\definecolor{darkgreen}{HTML}{089C25}
\newcommand{\newtext}[1]{#1}

\usepackage[normalem]{ulem}
\newcommand{\remove}[1]{}

\newcommand{\precrec}{\textsc{\small PrecRec}\xspace}
\newcommand{\preccorr}{\textsc{\small PrecRecCorr}\xspace}
\newcommand{\reverb}{\textsc{\small ReVerb}\xspace}
\newcommand{\restaurant}{\textsc{\small Restaurant}\xspace}
\newcommand{\book}{\textsc{\small Book}\xspace}
\newcommand{\synthetic}{\textsc{\small Synthetic}\xspace}

\newcommand{\union}{\textsc{\small Union}\xspace}
\newcommand{\cosine}{\textsc{\small Cosine}\xspace}
\newcommand{\estimate}{\textsc{\small 3-Estimate}\xspace}
\newcommand{\ltm}{\textsc{\small LTM}\xspace}
\newcommand{\solomon}{\textsc{\small Copy}\xspace}

\newcommand{\triple}[1]{\emph{\{#1\}}}
\newcommand{\true}[1]{#1}
\newcommand{\false}[1]{\neg #1}
\newcommand{\mS}{\mathcal{S}}
\newcommand{\mO}{\mathcal{O}}
\newcommand{\provide}[2]{#1 \models #2}
\newcommand{\notprovide}[2]{#1 \not\models #2}

\newcommand{\CPr}[2]{\Pr\left(#1 \mid #2\right)}
\newcommand{\CPrIndep}[2]{\Pr_{indep}\left(#1 \mid #2\right)}

\newcommand{\eat}[1]{}

\newcommand{\rbox}{\hfill $\Box$}
\newcommand{\ie}{{\em i.e.}}
\newcommand{\eg}{{\em e.g.}}
\newcommand{\etc}{{\em etc.}}

\makeatletter
\newcommand{\xRightarrow}[2][]{\ext@arrow 0359\Rightarrowfill@{#1}{#2}}
\makeatother

\newtheorem{definition}{Definition}[section]
\newtheorem{proposition}[definition]{Proposition}

\newtheorem{corollary}[definition]{Corollary}

\newtheorem{theorem}[definition]{Theorem}

\newtheorem{example}[definition]{Example}

\newcommand{\openbox}{\leavevmode
  \hbox to.77778em{%
  \hfil\vrule
  \vbox to.675em{\hrule width.6em\vfil\hrule}%
  \vrule\hfil}}
\newcommand{\qedsymbol}{\openbox}
\def\qed{\unskip\hspace{1em}\hspace*{\fill}\qedsymbol}
\makeatletter
\newcommand{\qedhere}{\hfill\qedsymbol\global\@qededtrue}
\makeatother

\newfont{\mycrnotice}{ptmr8t at 7pt}
\newfont{\myconfname}{ptmri8t at 7pt}
\let\crnotice\mycrnotice%
\let\confname\myconfname%

\permission{Permission to make digital or hard copies of all or part of this work for personal or classroom use is granted without fee provided that copies are not made or distributed for profit or commercial advantage and that copies bear this notice and the full citation on the first page. Copyrights for components of this work owned by others than the author(s) must be honored. Abstracting with credit is permitted. To copy otherwise, or republish, to post on servers or to redistribute to lists, requires prior specific permission and/or a fee. Request permissions from permissions@acm.org.}
\conferenceinfo{SIGMOD'14,}{June 22--27, 2014, Snowbird, UT, USA. \\
{\mycrnotice{Copyright is held by the owner/author(s). Publication rights licensed to ACM.}}}
\copyrightetc{ACM \the\acmcopyr}
\crdata{978-1-4503-2376-5/14/06\ ...\$15.00.\\
http://dx.doi.org/10.1145/2588555.2593674}

\usepackage{hyperref}
\hypersetup{%
  pdftitle = {Fusing Data with Correlations},
  pdfkeywords = {data fusion, correlation, data cleaning, data extraction},
  pdfauthor = {Ravali Pochampally, Anish Das Sarma, Xin Luna Dong, Alexandra Meliou, and Divesh Srivastava},
  bookmarksnumbered,
  bookmarksopen=false,
  urlcolor=black,
  linkcolor=black,
  citecolor=black,
  colorlinks=true,
  pdfstartview={FitH},
  pageanchor=false
}

\begin{document}


\title{Fusing Data with Correlations}

%

%

\numberofauthors{5} 

\author{
\alignauthor
\newtext{Ravali Pochampally\\
       \affaddr{University of Massachusetts}\\
       \email{\texttt{ravali@cs.umass.edu}}}
\alignauthor
\newtext{Anish Das Sarma\\
       \affaddr{Troo.Ly Inc.}\\
       \email{\texttt{anish.dassarma@gmail.com}}}
\alignauthor 
\newtext{Xin Luna Dong\\
       \affaddr{Google Inc.}\\
       \email{\texttt{lunadong@google.com}}}
\and  
\alignauthor 
\newtext{Alexandra Meliou\\
        \affaddr{University of Massachusetts}\\
        \email{\texttt{ameli@cs.umass.edu}}}
\alignauthor 
\newtext{Divesh Srivastava\\
        \affaddr{AT\&T Labs-Research}\\
        \email{\texttt{divesh@research.att.com}}}
}

\maketitle
\begin{abstract}
    Many applications rely on Web data and extraction systems to accomplish
    knowledge-driven tasks. Web information is not curated, so many sources
    provide inaccurate, or conflicting information. Moreover, extraction
    systems introduce additional noise to the data. We wish to automatically
    distinguish correct data and erroneous data for creating a cleaner set of
    integrated data. 
    Previous work has shown that a na\"ive voting strategy
    that trusts data provided by the majority or at least a certain number
    of sources may not work well in the presence of copying between the sources.
    However, correlation between sources can be much broader than copying: sources may provide
    data from complementary domains (\emph{negative correlation}), 
    extractors may focus on different types of information (\emph{negative correlation}), 
    and extractors may apply common rules in extraction (\emph{positive correlation, without copying}). 
    In this paper we present novel techniques modeling correlations between
    sources and applying it in truth finding. 
 \eat{Cleaning the data is not
    straightforward: extracted data is typically uncertain, and different
    sources often provide duplicate or conflicting results. Our framework
    automatically integrates results from multiple sources, targeting three
    main challenges: (a) Sources are generally black boxes. Internal details
    on how data is derived are unknown, and any integration scheme has to rely
    on high-level summaries of the source quality (e.g., precision and
    recall). (b) Sources may have unknown dependencies, but since access to
    their details is limited, these dependencies are hard to infer. (c) In
    several domains it is often possible, or necessary, for an entity to have
    multiple true values. In this paper, we present a technique for
    integrating data produced by multiple sources that addresses these three
    challenges.} We provide a comprehensive evaluation of our approach on
    three real-world datasets with different characteristics, as well as on synthetic data, 
    showing that our algorithms outperform the existing state-of-the-art techniques.
\end{abstract}

\renewcommand{\category}[3]{#1 [#2]: #3}
\noindent \textbf{Categories and Subject Descriptors:}\\

\renewcommand{\keywords}[1]{\noindent\textbf{Keywords:} #1}
\vspace{-0.2ex}

\section{Introduction}
\label{sec:introduction}
The Web is an incredibly rich source of information, which is growing at an unprecedented pace and is amassed by a plethora of contributors.  
An increasing number of users and applications rely on online data as the main resource to satisfy their information needs.  Web data is not curated and sources may often provide erroneous or conflicting information. 
Additionally, a lot of Web data is largely unstructured, lacking a predefined schema or consistent format. As a result, knowledge-driven applications in various domains (\eg, finance, technology, advertisement, \etc) rely on information extraction systems to retrieve structured relations from online sources.  However, extraction systems have less than perfect accuracy, invariably introducing more noise to the data.

Our goal is to automatically distinguish correct data and erroneous data for creating a cleaner set of data.
A na\"ive approach to achieve this goal is majority voting: we trust the data provided by the majority, or at least a certain number of sources. However, such a strategy may perform badly for two reasons. First, sources may provide data from complementary domains (\eg, information on scientific books vs. on biographies) and extractors may focus on different types of information (\eg, extracting from the Infobox or the texts of Wikipedia pages); blindly requiring agreement among sources may miss correct data and cause {\em false negatives}. Second, sources may easily copy and share data~\cite{BDD+09} and extractors may apply common rules; blindly trusting agreement among sources may enforce erroneous data and cause {\em false positives}. Such correlation or anti-correlation between sources makes it especially hard to tell the truth from wrong statements or extractions, as illustrated next.

\begin{example}
    \label{ex:extraction}
    Figure~\ref{fig:example} depicts example data extracted from the Wikipedia page for Barack Obama, using five different extraction systems.  Extracted data consist of \emph{knowledge triples} in the form of \triple{subject, predicate, object}; for example, \triple{Obama, spouse, Michelle} states that the spouse of {\em Obama} is {\em Michelle}. 
Some extracted triples are incorrect. For example, triple $t_2$ is false: extraction systems $S_1$ and $S_2$ derived the triple from a sentence referring to \emph{Barack Obama Sr}, rather than the current US president.  
 
Various types of correlations exist among the five sources. First, $S_1, S_4$ and $S_5$ 
implement similar extraction patterns and extract similar sets of triples; there is 
a {\em positive} correlation between these sources. Second, $S_3$ extracts triples from the Infobox of the
Wikipedia page while $S_1$ (similarly, $S_4$ and $S_5$) extracts triples from the text; 
their extracted triples are largely complementary and there is a {\em negative} correlation between them. 

Figure~\ref{fig:naiveVoting} shows the precision, recall, and F-measure of voting techniques:
Union-$k$ accepts a triple as true if at least $k\%$ of the extractors extract it; \eg, Union-25 accepts triples provided by at least 2 extractors:
it has high recall (missing only one triple), but makes a lot of mistakes 
(extracting 4 false triples) because of the common mistakes by the correlated 
sources $S_1$, $S_4$, and $S_5$.
Union-75 accepts triples provided by at least 4 extractors;
it misses a lot of true triples since $S_3$ is anti-correlated with three other sources. 
\end{example}

\begin{figure*}
\begin{minipage}[th]{\linewidth}
\centering
{\small
\begin{tabular}{|c|c|c|c|c|c|c|c|c|}
\hline
{\bf ID} & {\bf Web document} & {\bf KnowledgeTriple} & {\bf Correct?} & $\mathbf{S_1}$ & $\mathbf{S_2}$ & $\mathbf{S_3}$ & $\mathbf{S_4}$  & $\mathbf{S_5}$\\ 
\hline
$\mathbf{t_1}$ &  wiki/Barack\_Obama & \triple{Obama,profession,president} & Yes & \checkmark & \checkmark &  &   \checkmark  & \checkmark \\ 
\hline
$\mathbf{t_2}$ & wiki/Barack\_Obama & \triple{Obama,died,1982}  & No & \checkmark & \checkmark &   &  &  \\ 
\hline
$\mathbf{t_3}$ & wiki/Barack\_Obama &  \triple{Obama,profession,lawyer}  & Yes &  &  & \checkmark  &  &   \\ 
\hline
$\mathbf{t_4}$ & wiki/Barack\_Obama  & \triple{Obama,religion,Christian}  & Yes &   &  \checkmark & \checkmark & \checkmark  & \checkmark \\ 
\hline
$\mathbf{t_5}$ &  wiki/Barack\_Obama  & \triple{Obama,age,50}  & No &   &   \checkmark & \checkmark &  &   \\ \hline
$\mathbf{t_6}$ &  wiki/Barack\_Obama  &  \triple{Obama,support,White Sox} & Yes &  \checkmark &  &   & \checkmark & \checkmark  \\ 
\hline
$\mathbf{t_7}$ &  wiki/Barack\_Obama  &  \triple{Obama,spouse,Michelle}  & Yes &  \checkmark &  \checkmark & \checkmark   &  &   \\ 
\hline
$\mathbf{t_8}$ &  wiki/Barack\_Obama  & \triple{Obama,administered by,John G. Roberts}  & No & \checkmark  & \checkmark  &   & \checkmark  &  \checkmark  \\ 
\hline
$\mathbf{t_9}$ &  wiki/Barack\_Obama  &  \triple{Obama,surgical operation,05/01/2011} & No &  \checkmark & \checkmark  &   & \checkmark & \checkmark  \\ 
\hline
$\mathbf{t_{10}}$ &  wiki/Barack\_Obama  & \triple{Obama,profession,community organizer} & Yes & \checkmark  &    &  \checkmark & \checkmark & \checkmark   \\ 
\hline
\end{tabular}}
\vspace{-1mm}
\disable{
\subcaption{
Data extracted by five different extractors from the Wikipedia page for {\em Barack Obama}. The \checkmark\ symbols indicate which extraction systems produce each knowledge triple; for example, $t_3$ is extracted by $S_3$, but not by any other extractor. 
}
}
\label{fig:extractedData}
\end{minipage}%
    \vspace{-1mm}
    
\begin{minipage}[b]{.5\linewidth}
\centering
{\small
\begin{tabular}{|c|c|c|}
\multicolumn{1}{c}{} & \multicolumn{1}{c}{\phantom{precision}} & \multicolumn{1}{c}{\phantom{precision}}\\
\cline{2-3}
\multicolumn{1}{c|}{} & {\bf Precision} & {\bf Recall}  \\ \hline
$\mathbf{S_1}$ & $0.57$ & $0.67$ \\ \hline
$\mathbf{S_2}$ & $0.43$ & $0.5$ \\ \hline
$\mathbf{S_3}$ & $0.8$ & $0.67$ \\ \hline
$\mathbf{S_4}$ & $0.67$ & $0.67$ \\ \hline
$\mathbf{S_5}$ & $0.67$ & $0.67$ \\ \hline
\end{tabular}}
\hfill
{\small
\begin{tabular}{|c|c|c|}
\multicolumn{1}{c}{} & \multicolumn{1}{c}{\phantom{joint prec}} & \multicolumn{1}{c}{\phantom{joint rec}}\\
\cline{2-3}
\multicolumn{1}{c|}{} & {\bf Joint prec} & {\bf Joint rec}  \\ \hline
$\mathbf{S_2}\mathbf{S_3}$ & $0.67$ & $0.33$ \\ \hline
$\mathbf{S_1}\mathbf{S_3}$  & $1$ & $0.33$ \\ \hline
$\mathbf{S_1}\mathbf{S_2}\mathbf{S_4}$ & $0.33$ & $0.167$ \\ \hline
$\mathbf{S_1}\mathbf{S_4}\mathbf{S_5}$  & $0.6$ & $0.5$ \\ \hline
\end{tabular}}
\eat{\begin{tabular}{|c|c|c|c|c|}
    \multicolumn{4}{c}{}\\
\cline{2-5}
\multicolumn{1}{c|}{} & \multicolumn{2}{|c|}{\textbf{Correlation}} &  \multicolumn{2}{|c|}{\textbf{Type}} \\ 
\cline{2-5}
\multicolumn{1}{c|}{} & {\bf true t} & {\bf false t} &{\bf true t} & {\bf false t}\\
\hline
$\mathbf{S_2}\mathbf{S_3}$ & 1.0  & 1.0 & indep. & indep. \\ 
\hline
$\mathbf{S_1}\mathbf{S_3}$  & 0.75 & 0 & neg. & neg. \\ 
\hline
$\mathbf{S_1}\mathbf{S_2}\mathbf{S_4}$ & 0.75 & 1.33 & neg. & pos. \\
\hline
$\mathbf{S_1}\mathbf{S_4}\mathbf{S_5}$  & 1.69 & 2.67 & pos. & pos. \\ 
\hline
\end{tabular}}
\vspace{-1mm}
\disable{
\subcaption{
Precision and recall for each extractor, and joint precision and joint recall for some combinations of extractors.
}
}
\label{fig:methodAccuracy}
\end{minipage}%
\hfill
\begin{minipage}[b]{.45\linewidth}
\centering
{\small
\begin{tabular}{|c|c|c|c|}
\multicolumn{1}{c}{} & \multicolumn{1}{c}{\phantom{precision}} & \multicolumn{1}{c}{\phantom{precision}}
& \multicolumn{1}{c}{\phantom{F-measure}} \\
\cline{2-4}
 \multicolumn{1}{c|}{} & {\bf Precision} & {\bf Recall} & {\bf F-measure}  \\ \hline
\textbf{Union-25} & $0.56$ & $0.83$ & $0.67$ \\ \hline
\textbf{Union-50} & $0.71$ & $0.83$ & $0.77$ \\ \hline
\textbf{Union-75} & $0.6$ & $0.5$ & $0.55$ \\ \hline
\end{tabular}}
\vspace{-1mm}
\disable{
\subcaption{
Na\"ive fusion approaches based on voting do not achieve very good results, as they do not account for correlations among the extractors.
}
}
\label{fig:naiveVoting}
\end{minipage}
\vspace{-2mm}
\caption{Example~\ref{ex:extraction}: (a) knowledge triples derived by 5 extractors, (b) extractor quality and correlations, (c) voting results.
}\label{fig:example}
\vspace{-2mm}
\end{figure*}

\looseness -1
\emph{Data fusion} has studied resolving conflicts while considering source
copying~\cite{solomon2009, Dong2010vldb}. Previous approaches have two
limitations. First, they focus on copying of data between sources
and are based on the intuition that common mistakes are strong evidence of copying;
correlation is much broader: it can be positive or negative and 
can be caused by different reasons. Previous approaches are effective in detecting
positive correlation on false data, but are not effective with positive correlation on true data or negative correlation. Second, their model relies on the 
{\em single-truth} assumption such as everyone has a unique birthplace; 
however, in practice there can be multiple truths for certain ``facts'', such as someone may have multiple professions (\eg, triples $t_1$, $t_3$, and $t_{10}$ in Figure~\ref{fig:example} are all correct).

In this paper, we address the problem of finding truths among data provided by multiple sources, which may contain complex correlations. We make the following contributions.

\begin{itemize}[leftmargin=3mm, topsep=0mm, itemsep=-1mm]
    \item We propose measuring the quality of a source as its {\em precision}
    and {\em recall} and measuring the correlation between a subset of sources
    as their {\em joint precision} and {\em joint recall}. We express them in
    terms of conditional probability (Section~\ref{sec:overview}).
    \item We present a novel technique that derives the probability of a
    triple being true from the precision and recall of the sources using
    Bayesian analysis under the independence assumption
    (Section~\ref{sec:fusion_indep}). Our experiments show that even before
    incorporating correlations, our basic approach often outperforms existing
    state-of-the-art techniques.
    \item We extend our approach to handle correlations between the sources.
    We first present an {\em exact solution} that is exponential in the number
    of data sources. We then present two approximation schemes: the {\em
    aggressive approximation} reduces the computational complexity from
    exponential to linear, but sacrifices the accuracy of the predictions; our
    {\em elastic approximation} provides a mechanism to trade efficiency for
    accuracy (Section~\ref{sec:fusion_corr}).
\eat{    \item We propose effective approaches to derive the quality of sources and
    the correlations between them from a sample data set. Our technique
    compensates for low representation of certain types of answers (falses
    positives), as well as the presence of noise in the training set
    (Sections~\ref{sec:fusion_indep}-\ref{sec:fusion_corr}).
}
    \item We conduct a comprehensive evaluation of our techniques against
    three real-world data sets, as well as synthetic data. Our experiments
    show that our methods can significantly improve the results by considering
    correlation without adding too much overhead for efficiency
    (Section~\ref{sec:results}).
\end{itemize}

\section{The Fusion Problem}
\label{sec:overview} 
In this section, we introduce our data model and its semantics,  we provide a formal definition of the problem of fusing data from sources with unknown correlations, and we present a high-level overview of our approach. We summarize notations in Figure~\ref{tbl:notation}. 


\subsection{Data model}
\label{subsec:data_model}

We consider a set of data sources $\mS=\{S_1,\ldots,S_n\}$.
Each source provides some data and we call each unit of data a \emph{triple};
a triple can be considered as a cell in a database table in the form of
\triple{row-entity, column-attribute, value} (\eg, in a table about politicians,
a row can represent \emph{Obama}, a column can represent attribute \emph{profession},
and the corresponding cell can have value \emph{president}), or an RDF triple in the form
of \triple{subject, predicate, object}, such as \triple{Obama, profession, president}. 
We denote with $O_i$ the triples provided by source $S_i \in \mS$;
interchangeably, we denote with $\provide{S_i}{t}$ or $t \in O_i$ that $S_i$ provides triple $t$. 
Our data model consists of $\mS=\{S_1,\ldots,S_n\}$
and the collections of their output triples $\mO=\{O_1,\ldots,O_n\}$.  In a slight abuse of notation, we write $t\in\mO$ to denote that $\exists O_i\in\mO$ such that $t\in O_i$.  We use $\mO_t$ to represent the subset of outputs in $\mO$ that involve triple $t$; note that  $\mO_t$ contains the observation that a source $S_i$ does not provide $t$ only if $S_i$ provides other data in the domain of $t$, so we do not unnecessarily penalize data missing from irrelevant sources.

We consider deterministic sources: a source either outputs a triple, or it does not.  In practice, a source $S_i \in \cal S$ may provide a confidence score associated with each triple $t \in O_i$; we can consider that $S_i$ outputs $t$ if the assigned confidence score exceeds a certain threshold. 
As in previous work~\cite{solomon2009, ltm2012}, we assume that schema mapping and
reference reconciliation have been applied so we can compare the triples across sources.

Our goal is to purge the output of all incorrect triples to obtain a high-quality data set $R=\{t: t\in\mO\wedge t\ is\ true\}$.  We say that a triple $t$ is \emph{true} if it is consistent with the real world, and \emph{false} otherwise; for example, \triple{Obama, profession, president} is true whereas \triple{Obama, died, 1982} is false. We next show an instantiation of our data model for the data extraction scenario.

\begin{example}
Figure~\ref{fig:example} shows triples extracted by five extractors from the
{\em Wikipedia} page for {\em Barack Obama} and we need to determine which
triples are correctly extracted. We consider that each extractor corresponds to a
source; for example, $S_1$ corresponds to the first extractor and it provides
(among others) triple $t_1: \triple{Obama, profession, president}$. We denote
this as $\provide{S_1}{t_1}$, meaning that the extractor believes that $t_1$
is \remove{provided by} \newtext{a fact that appears on} the Wikipedia page. Accordingly, $O_1 = \{t_1, t_2, t_6, t_7,
t_8, t_9, t_{10}\}$.

\looseness -1
Based on $\mS$ and $\mO$, we decide whether each triple $t_i$ is true ($i \in [1,10]$). 
In this scenario, the extractor input (the processed web page)
represents the ``real world'', against which we evaluate the correctness of the
extractor outputs. We consider a triple to be true (\ie, correctly extracted) if the web
page indeed provides the triple.
\end{example}
\noindent
\textbf{Semantics:}
In this paper, we make two assumptions about semantics of the data:
First, we assume {\em triple independence}: the truthfulness of each triple is independent of that of other triples.
For example, whether the page indeed provides triple $t_1$ is independent of whether the page provides triple $t_2$.
Second, we assume \emph{open-world} semantics: a source considers any triple in its output as {\em true}, and any triple not in its output as \emph{unknown} (rather than {\em false}). For example, in Figure~\ref{fig:example}, $S_1$ provides $t_1$ and $t_2$ but not $t_3$, meaning that it considers $t_1$ and $t_2$ as being provided by the page, 
but does not know whether $t_3$ is also provided.
Note that this is in contrast with the {\em conflicting-triple, closed-world} semantics in~\cite{solomon2009};
under this semantics, \triple{Obama, religion, Christian} and \triple{Obama, religion, Muslim} would be considered
conflicting with each other, as we typically assume one can have at most one religion 
and a source claiming the former implicitly claims that the latter is false.

We make these assumptions for two reasons. The first reason is that they are suitable for many application scenarios.
One application scenario is data extraction, as shown in our motivating example:
when an extractor derives two different triples from a Web page
(often from different sentences or phrases), the correctness of the two extractions
are independent; if an extractor does not derive a triple from a Web page, it usually indicates that the extractor \emph{does not know} whether the page provides the triple, rather than that it \emph{believes} that the page does not provide the triple.
Another scenario is attributes that can accept multiple truths. For example, a person can have multiple professions:
the correctness of each profession is largely independent of other professions\footnote{\small Arguably, it is  unlikely for a person to be a doctor, a lawyer, and a plumber at
the same time as they require very different skills; 
we leave such joint reasoning with \emph{a priori} knowledge for future work.}, and a source that claims that
{\em Obama} is a president does not necessarily claim that {\em Obama} \emph{cannot} be a lawyer. 
The second reason is that, to the best of our knowledge, all previous work that studies correlation of sources 
focuses on the conflicting-triple, closed-world semantics; the independent-triple, open-world semantics allows us to 
fill the gap in the existing literature. Note that we can apply strategies for
conflicting-triple and closed-world semantics in the case of independent-triple
and closed-world semantics, or in the 
case of conflicting-triple and open-world semantics.
We leave combination of all semantics for future work. 

\subsection{Measuring truthfulness}\label{sec:probDef}
The objective of our framework is to distinguish true and false triples in a collection of source outputs. 
A key feature of our approach is that it does not assume any knowledge of the inner workings of the sources and how they derive the data that they provide.
First, this is indeed the case in practice for many real-world data sources --- they provide
the data without telling us how they obtain it.
Second, even when some information on the data derivation process is available, it may
be too complex to reason about; for example, an extractor often learns thousands (or even more) of patterns 
(e.g., distance supervision~\cite{mintz2009distant}) and uses internal coding to present them;
it is hard to understand all of them, let alone to reason about them and compare them across sources. 

Next, we show which key evidence we consider in our approach and then formally define our problem.



\subsubsection*{Source quality} 
\label{ssub:source_quality}

The quality of the sources affects our belief of the truthfulness of a triple. 
Intuitively, if a source $S$ has high {\em precision} (\ie, most of its provided triples
are true), then a triple provided by $S$ is more likely to be true.
On the other hand, if $S$ has a high {\em recall} (\ie, most of the true triples are provided
by $S$), then a triple not provided by $S$ is more likely to be false. 

We define precision and recall in the standard way: the precision $p_i$ of source $S_i \in \cal S$ represents the portion of triples in the output $O_i$ that are true; the recall $r_i$ of $S_i$ represents the portion of all true triples that appear in $O_i$. 
These metrics can be described in terms of probabilities as follows.
\begin{align}
    p_i &= \CPr{\true{t}}{\provide{S_i}{t}} \label{eq:precision}\\
    r_i &= \CPr{\provide{S_i}{t}}{\true{t}} \label{eq:recall}
\end{align}


\begin{example}\label{eg:sourceQuality}
Figure~\ref{fig:methodAccuracy} shows
the precision and recall of the five sources. For example, the precision of $S_1$ is $\frac{4}{7}=0.57$, as only 4 out of the 7 triples in $O_1$ are correct. The recall is ${4 \over 6} = 0.67$, as 4 out of the 6 correct triples are included in $O_1$. 
\end{example}

The recall of a source should be calculated with respect to the ``scope'' of its input.  
For example, if a source S provides only information about {\em Obama} but not about {\em Bush}, we may penalize the recall of S for providing only 1 out of the 3 professions of {\em Obama}, but should not penalize the recall of S for not providing any profession for {\em Bush}.
For simplicity of presentation, in the rest of the paper we ignore the ``scope'' of each source in our discussion, but all of our techniques work with either version of recall calculation.


\begin{figure}
\centering
{\small
\begin{tabular}{|l l|}
    \hline
\textbf{Notation} & \textbf{Description}\\
\hline
$\cal S$    & Set of sources $\mS=\{S_1,\ldots,S_n\}$\\
$O_i$       & Set of output triples of source $S_i$\\
$\mO$   	& $\mO=\{O_1,\ldots,O_n\}$\\
$\mO_t$   	& Subset of observations in $\mO$ that refer to triple $t$\\
$p_i$ (resp. $p_{\mS^*}$)       & Precision of source $S_i$ (resp. sources $\mS^*$)\\
$r_i$ (resp. $r_{\mS^*}$)     & Recall of source $S_i$ (resp. sources $\mS^*$)\\
$q_i$ (resp. $q_{\mS^*}$)      & False positive rate of $S_i$ (resp.  $\mS^*$)\\
$\provide{S_i}{t}$  & $S_i$ outputs $t$ ($t\in O_i$)\\
$\provide{\mS^*}{t}$  & $\forall S_i\in\mS^*$, $\provide{S_i}{t}$\\
$\CPr{\true{t}}{\mO}$      & Correctness probability of triple $t$ \\
\newtext{
$\Pr(\true{t})$, $\Pr(\false{t})$}   & \newtext{$\Pr(t=true)$ and $\Pr(t=false)$ respectively}
\\
\hline
\end{tabular}
}
\vspace{-.1in}
\caption{Summary of notations used in the paper.}
\label{tbl:notation}
\vspace{-2mm}
\end{figure}



\subsubsection*{Correlation}

Another key factor that can affect our belief of triple truthfulness
is the presence of correlations between data sources. Intuitively, if we know that two sources $S_i$ and $S_j$ are nearly
duplicates of each other, thus they are \emph{positively correlated}, the fact that both provide a triple $t$ should not significantly increase our belief that $t$ is true.
On the other hand, if we know two sources $S_i$ and $S_j$ are complementary and have little overlap,
so are \emph{negatively correlated}, the fact that a triple $t$ is provided by one but not the other
should not significantly reduce our belief that $t$ is true. 
Note the difference between \emph{correlation} and \emph{copying}~\cite{solomon2009}:
copying can be one reason for positive correlation, but positive correlation can 
be due to other factors, such as using similar extraction patterns or implementing
similar algorithms to derive data, rather than copying. 

We use \emph{joint precision} and \emph{joint recall} to capture correlation between sources.
The joint precision of sources in $\mS^*$, denoted by $p_{\mS^*}$, represents the portion of triples in the output of all sources in $\mS^*$ (\ie, intersection) that are correct; the joint recall of $\mS^*$, denoted by $r_{\mS^*}$, represents the portion of all correct triples that are output by all sources in $\mS^*$. 
If we denote by $\provide{\mS^*}{t}$ that a triple $t$ is output by all sources in $\mS^*$, 
we can describe these metrics in terms of probabilities as follows.
\begin{align}
    p_{\mS^*} &= \CPr{\true{t}}{\provide{\mS^*}{t}} \label{eq:precisionSet}\\
    r_{\mS^*} &= \CPr{\provide{\mS^*}{t}}{\true{t}} \label{eq:recallSet}
\end{align}



\begin{example}\label{ex:Corr}
Figure~\ref{fig:methodAccuracy} shows the joint precision and recall for selected subsets of sources.
Take the sources $\{S_1,S_4,S_5\}$ as an example. They provide similar sets of triples:
they all provide $t_1, t_6, t_8, t_9,$ and $t_{10}$. 
Their joint precision is ${3 \over 5} = 0.6$ and their joint recall is ${3 \over 6}=0.5$. 
Note that if the sources were independent, their joint recall would have been
$r_1\cdot r_4\cdot r_5 = 0.3$, much lower than the real one ($0.5$); this indicates {\em positive
correlation}.

On the other hand, $S_1$ and $S_3$ have little overlap in their data:
they both provide triples $t_7$ and $t_{10}$. 
Their joint precision is ${2 \over 2}=1$ and their joint recall is ${2 \over 6}=0.33$.
Note that if the sources were independent, their joint recall would have been
$r_1 \cdot r_3=0.45$, higher than the real one ($0.33$); this indicates {\em negative correlation}.

We define positive and negative correlation formally in Section~\ref{sec:fusion_corr}. 
\end{example}

\subsubsection*{Problem definition}
Our goal is to determine the truthfulness of each triple in $\mO$. We model the truthfulness of $t$ as the probability that $t$ is true, given the outputs of all sources;  we denote this as $\CPr{\true{t}}{\mO}$.
We can accept a triple $t$ as true if this probability is above 0.5, meaning that $t$ is more likely to be true than to be false.  
As we assume the truthfulness of each triple is independent, we can compute the probability for each triple separately conditioned on the provided data regarding $t$; that is, $\mO_t$.
We frame our problem statement based on source quality and correlation.  For now we assume the source quality metrics and correlation factors are given as input; we discuss techniques to derive them shortly.
We formally define the problem as follows:





\begin{definition}[Triple Truthfulness]
\label{defn:sap}
Given (1) a set of sources $\mS = \{S_1, \ldots, S_n\}$, (2) their outputs $\mO = \{O_1, \ldots, O_n\}$,  and (3) the joint precision $p_{\mS^*}$ and recall $r_{\mS^*}$ of each subset of sources $\mS^*\subseteq\mS$, compute the probability for each output triple $t\in\mO$, denoted by $\CPr{t}{\mO_t}$. 
\end{definition}  

\newtext{
Note that given a set $\mS$ of $n$ sources, there is a total of $2(2^n-1)$
joint precision and recall parameters. Since the input size is exponential in
the number of sources, even a polynomial algorithm will be infeasible in
practice. We show in Section~\ref{sec:fusion_corr} how we can reduce the
number of parameters we consider in our model and solve the problem
efficiently.
}

\subsection{Overview}
We start by studying the problem of triple probability computation under the assumption that sources are, indeed, independent.  We will show that even in this case, there are challenges to overcome in order to derive the probability.  We then extend our methods to account for correlations among sources.  Here, we present an overview of some high-level intuitions that we apply in each of these two settings.


\subsubsection*{Independent sources (Section~\ref{sec:fusion_indep})}

Fusion of data from multiple sources is challenging because the inner-workings of each source are not completely known.  We present a method that uses source quality metrics (precision and recall) to derive the probability that a source provides a particular triple, and applies Bayesian analysis to compute the truthfulness of each triple.  We describe how to derive the quality metrics if those are unknown. With this model, we are able to improve the F-measure to .86 (precision=.75, recall=1) for the motivating example.

\subsubsection*{Correlated sources (Section~\ref{sec:fusion_corr})}
Sources are often correlated: they may copy data from each other, employ similar techniques in deriving the data, or analyze complementary portions of the raw data sets.  Correlations can be positive or negative, and are generally unknown.
We address two main challenges in the case of correlated sources.


\begin{itemize}[leftmargin=3mm, topsep=0mm, itemsep=0mm]
\item \emph{Using correlations:}  We start by assuming that we know concrete correlations between sources. We will see that the main insight into revising the probability of triples is to determine how likely it is for a particular triple to have appeared in the output of a given subset of sources but not in the output of any other source.  Further, we use the inclusion-exclusion principle to express the correctness probability of a triple using the joint precision and joint recall of subsets of sources.

\item \emph{Exponential complexity:} The number of correlation parameters is exponential in the number of sources, which can make our computation infeasible. To counter this problem, we develop two approximation methods: our \emph{aggressive approximation} reduces the computation from exponential to linear, but sacrifices accuracy; our \emph{elastic approximation} provides a mechanism to trade efficiency for accuracy and improve the quality of our approximation incrementally. 

\eat{
\item \emph{Computing correlations:}  Since correlations are generally unknown, we describe a method to derive them.  Our technique is inspired by the Kappa measure~\cite{kappa} but incorporates the precision and recall of the sources as well.
}
\end{itemize} 
 
\noindent
Considering correlations, we can further improve the F-measure to 0.91 (precision=1, recall=0.83) for our motivating example, which is 18\% higher than Union-50 (\ie, majority voting).

\section{Fusing Independent Sources} 
\label{sec:fusion_indep}\label{SEC:FUSION_INDEP} 
In this section, we start with the assumption that the sources are independent.  Our goal is to estimate the probability that an output triple $t$ is true given the observed data: $\CPr{\true{t}}{\mO_t}$.
We describe a novel technique to derive this probability based on the quality of each source (Sec.~\ref{sec:probIndep}). Since these quality metrics are not always known in advance, we also describe how to derive them if we are given the ground truth on a subset of the extracted data (Sec.~\ref{sec:prec_rec}).

\subsection{Estimating triple probability }\label{sec:probIndep}
Given a collection of output triples for each source $O_i$, our objective is to compute, for each $t\in\mO$, the probability that $t$ is true, $\CPr{\true{t}}{\mO}$, based on the quality of each source.  
Due to the independent-triple assumption, $\CPr{\true{t}}{\mO} = \CPr{\true{t}}{\mO_t}$.

We use Bayes' rule to express $\CPr{\true{t}}{\mO_t}$ based on the inverse probabilities $\CPr{\mO_t}{\true{t}}$ and $\CPr{\mO_t}{\false{t}}$, which represent the probability of deriving the observed output data conditioned on $t$ being true or false respectively.  In addition, we denote the \emph{a priori} probability that $t$ is true with $\Pr(\true{t})=\alpha$.
\begin{align}
    \CPr{t}{\mO_t} = 
     \frac{\alpha\CPr{\mO_t}{\true{t}}}{\alpha\CPr{\mO_t}{\true{t}} +(1-\alpha)\CPr{\mO_t}{\false{t}}} \label{eq:bayesScore}
\end{align}
The denominator in the above expression is equal to $\Pr(\mO_t)$.
The a-priori probability $\alpha$ can be derived from a training set (\ie, a subset of the triples with known ground truth values, see Section~\ref{sec:prec_rec}).


We denote by $\mS_{t}$ the set of sources that provide $t$, and by $\mS_{\bar t}$ the set of sources that do not provide $t$.
\eat{\begin{align*}
    \mS_{t} = \{S_i\in\mS :\: \provide{S_i}{t}\}\\
    \mS_{\bar t} = \{S_i\in\mS :\: \notprovide{S_i}{t}\}  
\end{align*}
}Assuming that the sources are independent, the probabilities $\CPr{\mO_t}{\true{t}}$ and $\CPr{\mO_t}{\false{t}}$ can then be expressed using the \emph{true positive rate}, also known as \emph{sensitivity} or recall, and the \emph{false positive rate}, also known as the complement of \emph{specificity}, of each source as follows:
\begin{align}
    \CPr{\mO_t\!}{t} =\!\!\! \prod_{S_i\in\mS_t}\!\!\CPr{\provide{S_i}{t}}{\true{t}}\!\!\prod_{S_i\in\mS_{\bar t}}\!\!\left(1\!-\!\CPr{\provide{S_i}{t}}{\true{t}}\right)
    \label{eq:probIndep}
    \\
    \CPr{\mO_t\!}{\!\!\false{t}} \!=\!\!\! \prod_{S_i\in\mS_t}\!\!\!\CPr{\provide{S_i}{t}}{\!\false{t}}\!\!\!\prod_{S_i\in\mS_{\bar t}}\!\!\!\left(1\!-\!\CPr{\provide{S_i}{t}}{\!\false{t}}\right) \label{eq:probIndep2}
\end{align}

\eat{
\begin{align*}
    \CPr{\notprovide{S_i}{t}}{t}=1-r_i &&\text{and}&& \CPr{\notprovide{S_i}{t}}{\neg t}=1-q_i
\end{align*}

\begin{align}
    \CPr{\mO}{t} = \prod_{\substack{S_i\in\mS\\ \provide{S_i}{t}}}r_i\prod_{\substack{S_i\in\mS\\ \notprovide{S_i}{t}}}\left(1-r_i\right) \label{eq:probIndep}\\
    \CPr{\mO}{\neg t} = \prod_{\substack{S_i\in\mS\\ \provide{S_i}{t}}}q_i\prod_{\substack{S_i\in\mS\\ \notprovide{S_i}{t}}}\left(1-q_i\right)
\end{align}
}

From Eq.~\eqref{eq:recall}, we know $r_i=\CPr{\provide{S_i}{t}}{\true{t}}$. We denote the false positive rate by $q_i=\CPr{\provide{S_i}{t}}{\false{t}}$ and describe how we derive it in Section~\ref{sec:prec_rec}.
Applying these to Eq.~\eqref{eq:bayesScore}, we obtain the following theorem.
\begin{theorem}[Independent Sources]\label{thm:indep_sap}
Given a set of independent sources $\mS =\{S_1,\ldots, S_n\}$, the recall $r_i$ and the false positive rate $q_i$ of each source $S_i$, the correctness probability of an output triple $t$ is
$\CPr{\true{t}}{\mO_t}=\frac{1}{1+{1-\alpha \over \alpha}\cdot{1 \over \mu}}$,
where
\begin{align}
    \mu \;=\; \prod_{S_i\in\mS_t}\frac{r_i}{q_i}\prod_{S_i\in\mS_{\bar t}}\left(\frac{1-r_i}{1-q_i}\right)& \label{eq:indep}
\end{align}
\end{theorem}

Intuitively, we compute the correctness probability based on the (weighted) contributions of each source for each triple. Each source $S_i$ has contribution ${r_i \over q_i}$ for a triple that it provides,
and contribution ${1-r_i \over 1-q_i}$ for a triple that it does not provide. 
Given a triple $t$, we multiply the corresponding contributions of all sources to derive $\mu$, and then compute the probability of the triple accordingly.

We say a source $S_i$ is \emph{good} if it is more likely to provide a true triple than a false triple; that is, $\CPr{\provide{S_i}{t}}{\true{t}} > \CPr{\provide{S_i}{t}}{\false{t}}$ (\ie, $r_i > q_i$). Thus, a good source has a positive contribution for a provided triple --- once it provides a triple, the triple is more likely to be true; otherwise, the triple is more likely to be false. 
\begin{proposition}\label{prop:sourceQuality}
    Let $\mS'=\mS\cup\{S'\}$ and  $\mO'=\mO\cup\{O'\}$. 
\begin{itemize}[topsep=0mm, itemsep=0mm]
    \item If $S'$ is a good source:
    \begin{itemize}[topsep=0mm, itemsep=0mm]
        \item If $\provide{S'}{t}$, then $\CPr{\true{t}}{\mO'_t}>\CPr{\true{t}}{\mO_t}$.
        \item If $\notprovide{S'}{t}$, then $\CPr{\true{t}}{\mO'_t} < \CPr{\true{t}}{\mO_t}$.
    \end{itemize}
    \item If $S'$ is a bad source:
    \begin{itemize}[topsep=0mm, itemsep=0mm]
        \item If $\provide{S'}{t}$, then $\CPr{\true{t}}{\mO'_t}<\CPr{\true{t}}{\mO_t}$.
        \item If $\notprovide{S'}{t}$, then $\CPr{\true{t}}{\mO'_t} > \CPr{\true{t}}{\mO_t}$.
    \end{itemize}
\end{itemize}
\end{proposition}

\begin{example}\label{ex:indep}
    We apply Theorem~\ref{thm:indep_sap} to derive the probability of $t_2$, which is provided by $S_1$ and $S_2$ but not by $S_3, S_4,$ or $S_5$:
 \begin{align*}
        \mu= \frac{r_1}{q_1}\cdot \frac{r_2}{q_2} \cdot \frac{1-r_3}{1-q_3}\cdot \frac{1-r_4}{1-q_4}\cdot \frac{1-r_5}{1-q_5}
    \end{align*}
Suppose we know that $q_1=0.5$, $q_2=0.67$, $q_3=0.167$, and $q_4=q_5={0.33}$, and we know the recall as shown in Figure~\ref{fig:methodAccuracy}. Then we compute $\mu=0.1$. With $\alpha=0.5$, Theorem~\ref{thm:indep_sap} gives $\CPr{t_2}{\mO_{t_2}}=0.09$, so we correctly determine that $t_2$ is false.

However, assuming independence can lead to wrong results: $t_8$ is provided by $\{S_1,S_2,S_4,S_5\}$, but not by $S_3$. Using Eq.~\eqref{eq:indep} produces $\mu=1.6$ and $\CPr{t_8}{\mO_{t_8}}=0.62$, but $t_8$ is in reality false.
\end{example}

\eat{
According to Theorem~\ref{thm:indep_sap}, we can compute the probability for each triple in the data set in polynomial time as follows.
\begin{enumerate}[topsep=0mm, itemsep=0mm]
  \item For each source $S_i \in \cal S$, compute its vote count for a provided triple as $q_i \over r_i$ and its vote count for a non-provided triple as $1-q_i \over 1-r_i$. 
  \item For each triple $t \in \cup_iO_i$, compute its probability by Theorem~\ref{thm:indep_sap}. 
\end{enumerate}

\begin{proposition}
Let $n$ be the number of independent data sources in $\cal S$ and $m$ be the number of distinct triples in $\cal O$. We can compute the probability of each triple in time $O(mn)$. \rbox
\end{proposition}
}

%

 
\subsection{Estimating source quality}
\label{sec:prec_rec}
Theorem~\ref{thm:indep_sap} uses the recall and false positive rate of each source to derive the correctness probability. 
We next describe how we compute them from a set of training data, where we know the truthfulness of each triple.  
\newtext{
Existing work~\cite{Dong2012} also relies on training data to compute source
quality, while crowdsourcing platforms, such as Amazon Mechanical Turk,
greatly facilitate the labeling process~\cite{MW11}.
}

Computing the recall ($r_i$) relies on knowledge of the set of true triples, which is typically unknown {\em a priori}. Since we only need to decide truthfulness for each provided triple, we use the set of true triples that are provided by at least one source in the training data. Then, for each source $S_i, i \in [1,n],$ we count the number of true triples it provides and compute its recall according to the definition. In our motivating example (Figure~\ref{fig:example}), there are 6 true triples extracted by at least one extractor; accordingly, the recall of $S_1$ is ${4 \over 6}=0.67$ since it provides 4 true triples. 

However, we cannot compute the false positive rate ($q_i$) in a similar way by considering only false triples in the training data. We next illustrate the problem using an example.

\begin{example}
Consider deriving the quality of $S_1$ from the training set $\{t_1, \dots, t_{10}\}$.
Since the sources are all reasonably good, only 4 out of 10 triples are false.
If we compute $q_1$ directly from the data, we have $q_1={3 \over 4} = 0.75$. 
Since $q_1 > r_1=0.67$, we would (wrongly) consider $S_1$ as a bad source. 

Now suppose there is an additional source $S_0$ that provides 10 false triples $t_{11}-t_{20}$ and we include it in the training data. We would then compute $q_1 = {3 \over 14}=0.21$; suddenly, $S_1$ becomes a good source and much more trustable than it really is. 
\end{example}



To address this issue, we next describe a way that derives the false positive rate from the precision and recall of a source. The advantage of this approach is that the precision of a source can be easily computed according to the training data and would not be affected by the quality of other sources. Using Bayes' Rule on $\CPr{t}{\provide{S_i}{t}}$ we obtain a formula similar to Eq.~\eqref{eq:bayesScore}, and then we apply the conditional probability expressions for $p_i$, $r_i$, and $q_i$:
\begin{align*}
    &\CPr{t}{\provide{S_i}{t}} = 
     \frac{\alpha\CPr{\provide{S_i}{t}}{t}}{\alpha\CPr{\provide{S_i}{t}}{t} +(1-\alpha)\CPr{\provide{S_i}{t}}{\neg t}}
     \\
     &\xRightarrow{\eqref{eq:precision}\eqref{eq:recall}} p_i = \frac{\alpha r_i}{\alpha r_i + (1-\alpha)q_i}
     \Longrightarrow q_i=\frac{\alpha}{1-\alpha}\cdot\frac{1-p_i}{p_i}\cdot r_i
\end{align*}

For our example, we would compute the precision of $S$ as ${4 \over 7} = 0.57$.
Assuming $\alpha=0.5$, we can derive $q_1={0.5 \over 1-0.5}\cdot {1-0.57 \over 0.57}\cdot 0.67 = 0.5$, implying that $S_1$ is a borderline source, with fairly low quality (recall that $r_1=0.67>0.5$). 
Note that for $q_i$ to be valid, it needs to fall in the range of $[0,1]$.
The next theorem formally states the derivation and gives the condition for it to be valid. 

\begin{theorem}\label{thm:pqr}
Let $S_i, i \in [1,n],$ be a source with precision $p_i$ and recall $r_i$. 
\begin{itemize}[topsep=0mm, itemsep=0mm]
    \item If $\alpha \leq {p_i \over p_i+r_i-p_ir_i}$, we have $q_i=\frac{\alpha}{1-\alpha}\cdot\frac{1-p_i}{p_i}\cdot r_i$
    \item If $p_i > \alpha$, $S_i$ is a good source (\ie, $q_i < r_i$). 
\end{itemize}
\end{theorem}

Finally, we show in the next proposition that a triple provided by a high-precision source is 
more likely to be true, whereas a triple not provided by a \newtext{good,} high-recall source is more likely
to be false, which is consistent with our intuitions.

\begin{proposition}\label{prop:tripleProb}
    Let $\mS'=\mS\cup\{S'\}$ and $\mO'=\mO\cup\{O'\}$. 
    Let $\mS''=\mS\cup\{S''\}$ and $\mO''=\mO\cup\{O''\}$. 
The following hold. 
\begin{itemize}[topsep=0mm, itemsep=0mm]
    \item If $r_{S'}=r_{S''}$, $p_{S'} > p_{S''}$, and $\provide{S'}{t}$ and $\provide{S''}{t}$, then $\CPr{t}{\mO'_t} > \CPr{t}{\mO''_t}$.
    \item If $p_{S'}=p_{S''}$\newtext{$>\alpha$}, $r_{S'} > r_{S''}$, and $\notprovide{S'}{t}$ and $\notprovide{S''}{t}$, then $\CPr{t}{\mO'_t} < \CPr{t}{\mO''_t}$.
\end{itemize}
\end{proposition}

\subsection*{Comparison with LTM~\cite{ltm2012}}
The closest work to our independent model is the \emph{Latent Truth Model} 
(LTM)~\cite{ltm2012}; it treats source quality and triple correctness as 
latent variables and constructs a graphical model, and performs inference
using Gibbs sampling. LTM is similar to our approach in that (1) it also 
assumes triple independence and open-world semantics, and (2) its probability 
computation also relies on recall and false positive rate of each source.
However, there are three major differences. First, it derives the correctness
probability of a triple from the Beta distribution of the recall and false positive
rate of its providers; our model applies Bayesian analysis to maximize the
{\em a posteriori} probability. 
\newtext{ 
Using the Beta distribution enforces assumptions about the generative process
of the data, and when this model does not fit the actual dataset, \ltm has a
disadvantage against our non-parametric approach.
}
Second, it computes the recall 
and false positive rate of a source as the Beta distribution of 
the percentage of provided true triples and false triples; our model
derives false positive rate from precision and recall to avoid being biased
by very good sources or very bad sources. Third, it iteratively decides
truthfulness of the triples and quality of the sources; we derive source
quality from training data. 

We compare LTM with our basic approach experimentally,
showing that we have comparable results in general and sometimes better results; 
we also show that the correlation model we will present in the next section
obtains considerably better results than LTM, which assumes independence of sources.


\section{Fusing Correlated Sources} 
\label{sec:fusion_corr}\label{SEC:FUSION_CORR}

Theorems~\ref{thm:indep_sap} and~\ref{thm:pqr} summarize our approach for calculating the probability of a triple based on the precision and recall of each source.  In this section, we extend the results to account for correlations among sources. Before we proceed, we first show several scenarios where considering correlation between sources can significantly improve the results.

\begin{example}
Consider a set of $n$ good sources $\mS=\{S_1,\ldots,\\S_n\}$. All sources in $\mS$ have the same recall $r$ and false positive rate $q$, $r > q$. 
Given a triple $t$ provided by all sources, Theorem~\ref{thm:indep_sap}  computes $\mu_{indep}=({r \over q})^n$. 

\smallskip
\noindent
\textbf{Scenario 1 (Source copying):} Assume that all sources in $\mS$ are replicas. 
Ideally, we want to consider them as one source; indeed, their joint recall is
$r$ and joint false positive rate is $q$. Thus, we compute $\mu_{corr}={r
\over q}<\mu_{indep}$, which results in a lower probability for $t$; in other
words, a false triple would not get a high probability just because it is
copied multiple times.

\smallskip
\noindent
\textbf{Scenario 2 (Sources overlapping on true triples):} Assume that all sources in $\mS$ derive highly overlapping sets of true triples but each source makes independent mistakes (\eg, extractors 
that use different patterns to extract the same type of information). 
Accordingly, their joint recall is close to $r$ and their joint false positive rate is $q^n$. 
Thus, we compute $\mu_{corr} \approx {r \over q^n} > \mu_{indep}$, which results in a higher probability for $t$;
in other words, we will have much higher confidence for a triple provided by all sources.

\smallskip
\noindent
\textbf{Scenario 3 (Sources overlapping on false triples):} Consider the opposite case: all sources have a high overlap on false triples but each source provides true triples independently
(\eg, extractors that make the same kind of mistakes).
In this case, the joint recall is $r^n$ and the joint false positive rate is close to $q$. 
Thus, we compute $\mu_{corr} \approx {r^n \over q} < \mu_{indep}$, which results in a lower probability for $t$;
in other words, considering correlations results in a much lower 
confidence for a common mistake. 

\smallskip
\noindent
\textbf{Scenario 4 (Complementary sources):} Assume that all sources are
nearly complementary: their overlapping triples are rare but highly trustable 
(\eg, three extractors respectively focus on info-boxes, texts,
and tables that appear on a Wikipedia page). Accordingly, the sources have low joint recall
but very high joint precision; assume their joint recall is $r' \ll r$, and their joint false positive rate is $q'$, which is close to 0. 
Then, we compute $\mu_{corr} = {r' \over q'} \approx \infty$; in other words,
we highly trust the triples provided by all sources.

Under the same scenario, consider a triple $t'$ provided by only one source $S\in\mS$. 
Considering the negative correlation, the probability that a triple is provided only by $S$ is $r$ for true triples and $q$ for false triples;
thus, $\mu_{corr}'={r \over q} > {r \over q} \cdot ({1-r \over 1-q})^{n-1} = \mu_{indep}'$, which results in higher probability for $t'$. In other words, considering the negative correlation, the correctness probability of a triple won't be penalized if only a single source provides the triple.
\end{example}

These scenarios exemplify the differences of our work and copy detection in~\cite{Dong2010vldb}. Copy detection can handle scenario 1 appropriately; in scenarios 2 and 3 it may incorrectly conclude with copying and compute lower probability for true triples; it cannot handle anti-correlation in Scenario 4.

We first present an \emph{exact solution}, described by Theorem~\ref{thm:corr_sap}. However, exact computation is not feasible for problems involving a large number of sources, as the number of terms in the computation formula grows exponentially.  
In Section~\ref{sec:corr_linear}, we present an \emph{aggressive approximation}, which can be computed in linear time by enforcing several assumptions, but may have low accuracy.  
Our \emph{elastic approximation} (Section~\ref{sec:corr_ptime}) relaxes the assumptions gradually, and can achieve both good efficiency and good results.
Note that we can compute joint precision and joint recall, and derive
joint false positive rate exactly the same way as we compute them for a single source (Section~\ref{sec:prec_rec}).

\subsection{Exact solution} 
\label{sub:using_correlations}
Recall that Eq.~\eqref{eq:probIndep} and \eqref{eq:probIndep2} compute $\CPr{\mO_t}{t}$ and $\CPr{\mO_t}{\false{t}}$ by assuming independence between the sources. Now, we show how to compute them in the presence of correlations.
Using $\mS_{t}$ to represent the set of sources that provide $t$, and  $\mS_{\bar t}$ to represent the set of sources that do not provide $t$, we can express $\CPr{\mO_t}{t}$ as:
\begin{align}
        \CPr{\mO_t}{t} &= \Pr\!\!\left(\!\!\left(\bigwedge_{S \in \mS_t}\!\provide{S}{t}\right) \wedge \left(\bigwedge_{S' \in \mS_{\bar t}}\! \notprovide{S'}{t}\right)\biggm\vert t\right)
     \label{eq:step1}
\end{align}
\eat{ The event $(t\not\models\mS_{\bar t})$ is mutually exclusive with all events $(\provide{S_i}{t})$, where $S_i\in\mS_{\bar t}$:  A triple $t$ either is not outputted by any source in $\mS_{\bar t}$, or it is output by at least one source $S_i\in\mS_{\bar t}$.
Based on this observation, we rewrite Eq.~\eqref{eq:step1}:
\begin{align}
        \CPr{\mO_t}{t} =& \CPr{t\models\mS_{t}}{t} \nonumber\\ 
        &- \CPr{\left(t\models\mS_{t}\right)\cap\left(\bigcup_{S_i\in\mS_{\bar t}}\provide{S_i}{t}\right)}{t} \label{eq:step2}
\end{align} }

We apply the \emph{inclusion-exclusion principle} to rewrite the formula using the joint recall of the sources: 

\begin{align}
    \CPr{\mO_t}{t} &= \sum_{\mS^*\subseteq\mS_{\bar t}} \left(-1\right)^{|\mS^*|}\CPr{\provide{\{\mS_{t}\cup\mS^*\}}{t}}{t} \nonumber\\ 
    &= \sum_{\mS^*\subseteq\mS_{\bar t}} \left(-1\right)^{|\mS^*|} r_{\mS_{t}\cup\mS^*} 
\label{eq:step3} 
\end{align}

Note that when the sources are independent, Eq.~\eqref{eq:step3} computes exactly
$\prod_{S_i\in\mS_t}r_i\prod_{S_i\in\mS_{\bar t}}(1-r_i)$, which is equivalent to Eq.~\eqref{eq:probIndep}.
We compute $\CPr{\mO}{\false{t}}$ in a similar way, using the joint false positive rate of the sources, which can be derived from joint precision and joint recall as we described in Theorem~\ref{thm:pqr}:

\begin{align}
    \CPr{\mO_t}{\false{t}} = \sum_{\mS^*\subseteq\mS_{\bar t}} \left(-1\right)^{|\mS^*|} q_{\mS_{t}\cup\mS^*} \label{eq:step4} 
\end{align}

Theorem~\ref{thm:corr_sap} extends Theorem~\ref{thm:indep_sap} for the case of correlated sources.
\begin{theorem}\label{thm:corr_sap}
Given a set of sources $\mS = \{S_1, \ldots, S_n\}$, the joint recall and joint false positive rate for each subset of the sources, the probability of a triple $t$ is
$\CPr{t}{\mO}=\frac{1}{1+{1-\alpha \over \alpha}\cdot{1 \over \mu}}$,
where
\begin{align}
    \mu \;=\; \frac{\CPr{\mO_t}{t}}{\CPr{\mO_t}{\false{t}}}
\end{align}
and $\CPr{\mO_t}{t}$, $\CPr{\mO_t}{\false{t}}$ are computed by Eq.~\eqref{eq:step3} and~\eqref{eq:step4}. 
\end{theorem}

\begin{corollary}
Given a set $\mS = \{S_1, \ldots, S_n\}$, where all sources are independent, the correctness probabilities computed using Theorems~\ref{thm:indep_sap} and~\ref{thm:corr_sap} are equal. 
\end{corollary}

\begin{example}\label{ex:correlation}
Triple $t_8$ of Figure~\ref{fig:extractedData} is provided by $\mS_{t_8}=\{S_1,S_2,S_4,S_5\}$. 
We use notations $r_{\{S_1, S_2, S_4, S_5\}}$ and $r_{1245}$ interchangeably. We can compute joint recall for a set of sources as we do for a single source (Section~\ref{sec:prec_rec}), but here we assume that all the joint recall and joint false positive rate parameters are given. 

We compute $\CPr{\mO_t}{t_8}$ and $\CPr{\mO_t}{\false{t_8}}$, according to Eq.~\eqref{eq:step3}:
    \begin{align*}
        \CPr{\mO_{t_8}}{t_8} = &r_{1245} - r_{12345} = 0.22 - 0.11 = 0.11\\
		\CPr{\mO_{t_8}}{\false{t_8}} = &q_{1245} - q_{12345} = 0.22 - 0.037 = 0.185
    \end{align*}
Assuming a-priori probability $\alpha=0.5$, we derive $\CPr{t_8}{\mO} = {1 \over 1 + {0.185 \over 0.11}} = 0.37$.
Note that although $t_8$ is provided by four out of the five sources, $S_1, S_4$, and $S_5$ are correlated, which reduces their contribution to the correctness probability of $t_8$. Using correlations allows us to correctly classify $t_8$ as false, whereas the independence assumption leads to the wrong result, as shown in Example~\ref{ex:indep}.
\end{example}

Even though accounting for correlations can significantly improve accuracy, it increases the computational cost. The computation of $\CPr{\mO_t}{t}$ and $\CPr{\mO_t}{\false{t}}$ is exponential in the number of sources that do not provide $t$, thus impractical when we have a large number of sources. We next describe two ways to approximate $\CPr{\mO_t}{t}$ and $\CPr{\mO_t}{\false{t}}$.

\subsection{Aggressive approximation}
\label{sec:corr_linear}
In this section, we present a linear approximation that reduces the total number of terms 
in the computation by enforcing a set of assumptions. 
We first present the main result for the approximation in Definition~\ref{thm:corr_linear}, and we show how we derive it later. 
\begin{definition}[Aggressive approximation]\label{thm:corr_linear}
Given a set of sources $\mS = \{S_1, \ldots, S_n\}$, \newtext{the recall $r_i$ and false positive rate $q_i$ of each source $S_i$,} and the joint recall and joint false positive rate for \remove{each} \remove{sub}set\newtext{s $\mS$ and $\mS-S_i$}, the \emph{aggressive approximation} of the probability $\CPr{t}{\mO_t}$ is defined as:
$\frac{1}{1+{1- \alpha \over \alpha}\cdot{1 \over \mu_{aggr}}}$,
where
\begin{align}
    \mu_{aggr} &= \prod_{S_i\in\mS_t}\frac{C_i^+r_i}{C_i^- q_i}\prod_{S_i\in\mS_{\bar t}}\left(\frac{1-C_i^+r_i}{1-C_i^- q_i}\right)& \label{eq:corr} \\
    C_i^+ &= {r_{1\dots n} \over r_i\cdot r_{12\dots(i-1)(i+1)\dots n}} \label{eq:c_iplus}\\
    C_i^- &= {q_{1\dots n} \over q_i\cdot q_{12\dots(i-1)(i+1)\dots n}}\label{eq:c_iminus}
\end{align}
\end{definition}

Eq.~\eqref{eq:corr} differs from~\eqref{eq:indep} in that it replaces $r_i$ (resp. $q_i$) with $C_i^+r_i$ (resp. $C_i^-q_i$). 
Intuitively, $C_i^+$ and $C_i^-$ represent the correlation between $S_i$ 
and the rest of $\mS$, in the case of true and false triples respectively. Eq.~\eqref{eq:corr} weighs $r_i$ and $q_i$ by these ``correlation'' parameters. 
When the sources are independent, $C_i^+=C_i^-=1$, and the approximation obtains the same result as Theorem~\ref{thm:indep_sap}.
\newtext{
In contrast with Definition~\ref{defn:sap}, aggressive approximation only uses $2n+1$ instead of $2(2^n-n-1)$ correlation parameters.
}

\begin{figure}[t]
\centering
{\small
\begin{tabular}{cccccc}
\toprule
 & $S_1$ & $S_2$ & $S_3$ & $S_4$ & $S_5$ \\
\midrule
$C^+$ & ${0.11 \over 0.67 * 0.167} = 1$ & 1 & 0.75 & 1.5 & 1.5 \\
$C^-$ & ${0.037 \over 0.5 * 0.037} = 2$ & 1 & 1 & 3 & 3 \\
\bottomrule
\end{tabular}
}
\vspace{-1mm}
\caption{Correlation parameters of the aggressive approximation computed for each source of Figure~\ref{fig:extractedData}.
\label{tbl:corrPara}}
\vspace{-.1in}
\end{figure}

\begin{corollary}\label{cor:indepEquiv}
Given a set $\mS = \{S_1, \ldots, S_n\}$, where all sources are independent, the correctness probabilities computed using Theorem~\ref{thm:indep_sap} and Definition~\ref{thm:corr_linear} are  the same.
\end{corollary}
\begin{example}
Consider triple $t_8$ in Figure~\ref{fig:extractedData}. 
Figure~\ref{tbl:corrPara} shows the correlation parameters for each source and illustrates how they are computed for $S_1$. 
These parameters indicate that $S_1$, $S_4$ and $S_5$ are positively correlated for false triples; for true triples, $S_3$ is anti-correlated with the rest of the sources, whereas $S_4$ and $S_5$ are correlated.
Accordingly, we compute $\mu_{aggr}$ as follows:
 \begin{align*}
        \mu_{aggr} &= \frac{0.67\cdot 0.5\cdot(1- 0.75\cdot0.67)\cdot1.5\cdot0.67\cdot 1.5\cdot0.67}{2\cdot 0.5\cdot0.67\cdot (1- 0.167)\cdot 3\cdot 0.33 \cdot3\cdot 0.33}=0.3
    \end{align*}

Thus, we compute $\CPr{t_8}{\mO}=\frac{1}{1+\frac{1}{\mu_{aggr}}}=0.23$, which is lower than the exact computation in Example~\ref{ex:correlation}.  Both approaches correctly determine that $t_8$ is false. 
\end{example}

Obviously, the computation is linear in the number of sources. Also, instead of having an
exponential number of joint recall and false positive rate values, we only need $C_i^+$
and $C_i^-$ for each $S_i, i \in [1,n]$, which can be derived from a linear
number of joint recall and false positive rate values. However, as the following proposition
shows, this aggressive approach can produce bad results for special cases with
strong correlation (\ie, sources are replicas), or strong
anti-correlation (\ie, sources are complementary to each other).

\begin{proposition}\label{prop:aggrBadCases}
If all sources in $\mS$ provide the same data, Definition~\ref{thm:corr_linear} computes probability $\alpha$ for each provided triple.

If every pair of sources in $\mS$ are complementary to each other. Definition~\ref{thm:corr_linear} does not compute a valid probability for any triple.
\end{proposition}


Next, we proceed to describe the three major steps that lead to Definition~\ref{thm:corr_linear}.

\subsubsection*{I. Correlation factors}
Accounting for correlations, the probability that a set ${\mS}^*$ of
sources all provide a true triple is $r_{{\mS}^*}$ instead of
$\prod_{S_i\in\mS^*}r_i$ (similarly for a false triple). 
We define two \emph{correlation factors}:
$C_{\mS^*}$ and $C_{\mS^*}^\neg$:
\begin{align}
    C_{\mS^*} &= \frac{\CPr{\provide{\mS^*}{t}}{t}} {\CPrIndep{\provide{\mS^*}{t}}{t}}
= \frac{r_{{\mS}^*}}{\prod_{\substack{S_i\in\mS^*}}r_i} \label{eq:corrFactorPos}
    \\
        C_{\mS^*}^\neg &= \frac{\CPr{\provide{\mS^*}{t}}{\false{t}}} {\CPrIndep{\provide{\mS^*}{t}}{\false{t}}} 
= \frac{q_{{\mS}^*}}{\prod_{\substack{S_i\in\mS^*}}q_i}  \label{eq:corrFactorNeg}
\end{align}
If the sources in $\mS^*$ are independent, then $C_{\mS^*}=C_{\mS^*}^\neg=1$.  Deviation from independence may produce values greater than 1, which imply {\em positive correlations} (\eg, for $S_4$ and $S_5$ in Figure~\ref{fig:extractedData}, $C_{45}={0.67 \over 0.67\cdot 0.67}=1.5>1$), or lower than 1, which imply {\em negative correlations}, also known as {\em anti-correlations} (\eg, for $S_1, S_3$ in Figure~\ref{fig:extractedData}, $C_{13}={0.33 \over 0.67\cdot 0.67}=0.75<1$). 

Using separate parameters for true triples and false triples allows for a richer representation of correlations.  In fact, two sources may be correlated differently for true and false triples.  
For example, sources $S_2$ and $S_3$ in Figure~\ref{fig:extractedData} are independent with respect to true triples ($C_{23}=1$), and negatively correlated with respect to false triples ($C_{23}^\neg=0.5<1$).
%

Using correlation factors, $\CPr{\mO}{t_8}$ in our running example can be rewritten as follows:
    \begin{align*}
        \CPr{\mO_{t_8}}{t_8} =& C_{1245}r_1r_2r_4r_5 - C_{12345}r_1r_2r_3r_4r_5    
    \end{align*}

\subsubsection*{II. Assumptions on correlation factors}
To transform the equations with correlation factors into a simpler form,
we make \emph{partial independence assumptions}. Before we formally state the assumptions, 
we first illustrate it using an example.

\begin{example}\label{ex:approximation}
Consider sources ${\mS}=\{S_1\ldots S_5\}$ and
assume $S_4$ is independent of the set of sources $\{S_1, S_2, S_3\}$ and of sources $\{S_1, S_2, S_3, S_5\}$. Then, we have 
$r_{123}\cdot r_{4} = r_{1234}$ and $r_{12345} = r_{1235}\cdot r_{4}$.
Thus, $r_{123}r_{12345} = r_{1234}r_{1235}$.
Using the definition of the correlation factors, it follows that 
$C_{123}C_{12345} = C_{1234}C_{1235}$. 

Accordingly, we can rewrite the correlation factors;
for example $C_{123}={C_{1234}C_{1235} \over C_{12345}}$, and similarly,
$C_{23}={C_{1234}C_{1235}C_{2345} \over (C_{12345})^2}$.
Combining  Eqs~\eqref{eq:c_iplus} and \eqref{eq:corrFactorPos}, the following equations hold: $C_1^+={C_{12345} \over C_{2345}}$, $C_4^+ = {C_{12345} \over C_{1235}}$, $C_5^+={C_{12345} \over C_{1234}}$.  Using these, we can rewrite the following correlation factors:
$C_{123}={C_{12345} \over C_4^+C_5^+}$ and $C_{23}={C_{12345} \over C_1^+C_4^+C_5^+}$. 
\end{example}

According to Eqs.(\ref{eq:c_iplus}--\ref{eq:corrFactorNeg}), we can compute
$C_i^+={C_{\mS} \over C_{{\mS} \setminus \{S_i\}}}$ and
$C_i^-={C_{\mS}^\neg \over C_{{\mS} \setminus \{S_i\}}^\neg}$.
As illustrated in Example~\ref{ex:approximation}, partial independence assumptions lead to the following equations:
    \begin{align}
C_{{\mS}^*} = {C_{\mS} \over \prod_{S_i\in\mS\setminus{\mS}^*} C_i^+} && \text{ and } &&
C_{{\mS}^*}^\neg = {C_{\mS}^\neg \over \prod_{S_i\in\mS\setminus{\mS}^*} C_i^-} 
    \end{align}
As a special form, when ${\mS}^*=\emptyset$, we have $C_{{\mS}^*}=C_{{\mS}^*}^\neg=1$, so,
\begin{align}
 C_{\mS} = \prod_{S_i\in\mS} C_i^+ && \text{ and } && 
 C_{\mS}^\neg = \prod_{S_i\in\mS} C_i^- \label{eq:all+} 
\end{align}
Under these assumptions, $\CPr{\mO}{t_8}$ in our running example can be rewritten as follows:
    \begin{align*}
        \CPr{\mO_{t_8}}{t_8} =& {C_{12345} \over C_3^+}r_1r_2r_4r_5 - C_{12345}r_1r_2r_3r_4r_5    
    \end{align*}

\subsubsection*{III. Transformation}
We are ready to transform the equation into a simpler form, which is the same as the one in Definition~\ref{thm:corr_linear}. We continue illustrating the main intuition with our running example:
    \begin{align}
        \CPr{\mO_{t_8}}{t_8} &= {C_{12345} \over C_3^+}r_1r_2r_4r_5(1 - C_3^+r_3)\nonumber\\
        \overset{\eqref{eq:all+}}{=\joinrel=} &{C_1^+C_2^+C_3^+C_4^+C_5^+ \over C_3^+}r_1r_2r_4r_5(1 - C_3^+r_3)\nonumber\\
        =& (C_1^+r_1)(C_2^+r_2)(1-C_3^+r_3)(C_4^+r_4)(C_5^+r_5)\label{eq:t8aggressive}
    \end{align}

\subsection{Elastic approximation}
\label{sec:corr_ptime}
So far we have presented two solutions: the exact solution gives precise probabilities but is
computationally expensive; the aggressive approximation enforces partial independence assumptions
resulting in linear complexity, but in the worst case can compute probabilities independent of the quality of the sources. 
In this section, we present an elastic approximation algorithm
that makes a tradeoff between efficiency and quality.

The key idea of the elastic approximation is to use the linear approximation as a starting point 
and gradually adjust the results by relaxing the assumptions in every step.
We call the algorithm ``elastic'' because it can be configured to iterate over different levels of adjustments, depending on the desired level of approximation.  
We illustrate this idea with our running example.

\begin{example}
Triple $t_{8}$ is provided by four sources $\mS_{t_8}=\{S_1,S_2,S_4,S_5\}$ (Figure~\ref{fig:extractedData}).  We will adjust the linear approximation of $\CPr{\mO_{t_8}}{t_8}$ from Eq.~\eqref{eq:t8aggressive}, by adding specific terms at every level. We refer to the \emph{degree} of a term in the aggressive approximation, as the number of recall (or false positive rate) parameters associated with that term. The aggressive approximation for $\CPr{\mO_{t_8}}{t_8}$ contains two terms of degrees 4 and 5: $C_1^+C_2^+C_4^+C_5^+r_1r_2r_4r_5$ and $C_1^+C_2^+C_3^+C_4^+C_5^+r_1r_2r_3r_4r_5$ respectively (directly derived from Eq.~\eqref{eq:t8aggressive}).

\looseness -1
Elastic approximation makes corrections to the aggressive approximation based on terms of a given degree at every level.
At level-0 we consider the terms with degree of $|{\mS}_{t_8}|+0=4$, \ie, the term $C_1^+C_2^+C_4^+C_5^+r_1r_2r_4r_5$;
the exact coefficient of the term is $C_{1245}$ but we approximated it to $C_1^+C_2^+C_4^+C_5^+$
based on the assumption that $C_{1245} = {C_{12345} \over C_3^+} = C_1^+C_2^+C_4^+C_5^+$. 
To remove the assumption, we need to replace $(C_1^+r_1)(C_2^+r_2)(C_4^+r_4)(C_5^+r_5)$ with $C_{1245}r_1r_2r_4r_5=r_{1245}$. 
Since $r_{1245}=q_{1245}=0.22$, we have 
 \begin{align*}
        \mu &= \frac{0.22}{0.22}\cdot \frac{1-0.75\cdot0.67}{1-0.167}= 0.6
    \end{align*}
Note that the level-0 adjustment affects not only terms with degree 4, but actually all terms as we show next.

At level-1, we consider the terms with degree of $|{\mS}_{t_8}|+1=5$.  After  level-0 adjustment, the 5-degree term is $C_{1245}C_3^+r_1r_2r_3r_4r_5$.  We will replace $C_{1245}C_3^+$ with the exact coefficient $C_{12345}$, which will now give us the exact solution.

In summary, the $\mu_{aggr}$ parameter calculated by the aggressive approximation, the level-0 adjustment, and the level-1 adjustment are 0.3, 0.6, and 0.59 respectively.
Note that, as is the case in this example, we don't need to compute all the levels; stopping after a constant number of levels can get close to the exact solution.
\end{example}

\begin{algorithm}[t]
\caption{\textsc{Elastic} (Elastic approximation)}
\label{alg:ptime_corr}
{\small
\begin{algorithmic}[1]
\algtext*{EndFor}
\State $R \gets r_{\mS_{t}} \displaystyle \prod_{S_i \in {\mS}_{\bar t}}(1-C_i^+r_i)$; \label{ln:r}
\State $Q \gets q_{\mS_{t}} \displaystyle \prod_{S_i \in {\mS}_{\bar t}}(1-C_i^-q_i)$; \label{ln:q}
\For{$l=1 \to \lambda$} \label{ln:level}\textcolor{gray}{\Comment{$\lambda \geq 1$ is the desired adjustment level}}
\ForAll {subsets $\mS^* \subseteq {\mS}_{\bar t}$ of size $l$} \label{ln:term}
\State $\mS_l \gets \{\mS_{t} \cup \mS^*\}$;\label{ln:term2}
\State $R \gets R + (-1)^l ( C_{\mS_l}  - C_{\mS_t}\prod_{S_i \in {\mS}^*}C_i^+) \prod_{S_i \in {\mS}_l} r_i$; \label{ln:rdiff}
\State $Q \gets Q + (-1)^l ( C_{\mS_l}^\neg  - C_{\mS_t}^\neg\prod_{S_i \in {\mS}^*}C_i^-) \prod_{S_i \in {\mS}_l} q_i$;\label{ln:qdiff}
\EndFor 
\EndFor
\State\Return $R \over Q$;\label{ln:return} 
\end{algorithmic}
}
\end{algorithm}

Our {\sc Elastic} algorithm (Algorithm~\ref{alg:ptime_corr}) contains the pseudo code of our elastic approximation. 
Lines~\ref{ln:r}--\ref{ln:q} compute the initial values of the numerator $R$ and denominator $Q$
for $\mu$. Note that they have already applied the level-0 adjustment. 
Then for each level $l$ from 1 up to the required level $\lambda$ (line~\ref{ln:level}), 
we consider each term with degree $|{\mS}_t|+l$ (lines~\ref{ln:term}--\ref{ln:term2}), and make up the 
difference between the exact coefficient and the approximate coefficient
(lines~\ref{ln:rdiff}--\ref{ln:qdiff}). Finally, line~\ref{ln:return} returns $R \over Q$ 
as the value of $\mu$.

\begin{proposition}\label{prop:runtime}
Given a set of $n$ sources, a set of $m$ triples for probability computation,
and an approximation level $\lambda$, {\sc Elastic}
takes times $O(m\cdot n^\lambda)$ and the number of required correlation parameters
is in $O(m\cdot n^\lambda)$. 
\end{proposition}

\section{Evaluation} 
\label{sec:results}
This section describes a thorough evaluation of our models on three real-world
datasets as well as synthetic data. Our experimental results show that 
(1) considering correlations between sources can significantly improve fusion results;
(2) our elastic approximation can effectively estimate triple probability
with much shorter execution time; and (3) even in presence of only independent
sources, our model can outperform state-of-the-art data fusion approaches.


\eat{
In this section, we perform a thorough evaluation of our main fusion methods:
\emph{\precrec}, which estimates triple probability scores using an
independence assumption as in Theorem~\ref{thm:indep_sap}, and
\emph{\preccorr}, which models correlations as in Theorem~\ref{thm:corr_sap}.
We further analyze our aggressive and elastic approximations for \preccorr,
described in Definition~\ref{thm:corr_linear} and
Algorithm~\ref{alg:ptime_corr} respectively.

We conducted our evaluation using three real-world datasets, as well as synthetic data, and we compared our techniques with several baselines and other state-of-the-art approaches.  }

\subsubsection*{Datasets}
We first describe the real-world datasets we used in our experiments; we describe our synthetic data generation in Section~\ref{sec:synthetic}.

\vspace{1mm}
\noindent
\textbf{\reverb:} The ReVerb ClueWeb Extraction
dataset~\cite{Fader11} samples
500 sentences from the Web using Yahoo's random link
service and uses 6 extractors to extract triples from these sentences.
The gold standard contains 2407 extracted triples 
(616 true and 1791 false).

\vspace{1mm}
\noindent
\textbf{\restaurant:} The restaurant dataset from~\cite{MW11} consists of triples on the location of a collection of 1000 restaurants provided by 7 sources ({\em Yelp, Foursquare, OpenTable, MechanicalTurk,
YellowPages, CitySearch, MenuPages}). 
The gold standard contains 93 triples 
(68 true and 25 false), selected by majority vote over 10 Mechanical Turk responses.

\vspace{1mm}
\noindent
\textbf{\book:} The book dataset from~\cite{solomon2009} was collected by crawling {\em abebooks.com}.
The dataset consists of 5900 unique book-author triples from 879 seller sources. The
gold standard consists of 225 randomly sampled books for which the authors are
manually identified from book covers; 482 authors are correctly provided for these books
and 935 authors are wrongly provided. Note that our version of this dataset has more noise than the
one used in~\cite{ltm2012}, resulting in a more challenging setting.

\begin{wrapfigure}[12]{l}{0.2\textwidth}
    \centering
        \includegraphics[scale=0.65]{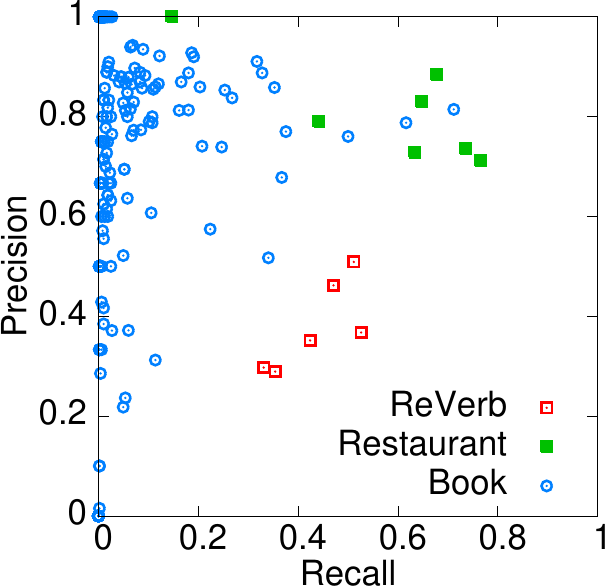}
    \label{fig:figs_sourceQuality}
\end{wrapfigure}
We observe that these datasets display varied characteristics:
the sources in \restaurant all have high precision, and most have high recall;
the sources in \reverb have fairly low precision and recall;
the sources in \book have large variations in precision, and most of them have low recall.
Such differences allow us to evaluate our models in a variety of scenarios.

\eat{
\vspace{1mm}
\noindent
\textbf{\synthetic:} We generate synthetic data to test our methods under different correlation scenarios and to evaluate the runtime-accuracy trade-off of our approaches.  We vary the quality of sources, the number of sources, the size of the dataset, and the types and strength of correlations among sources.
}

\disable{
\begin{figure*}[t]
\begin{subfigure}[b]{\linewidth}
\centering
\includegraphics[scale=0.7]{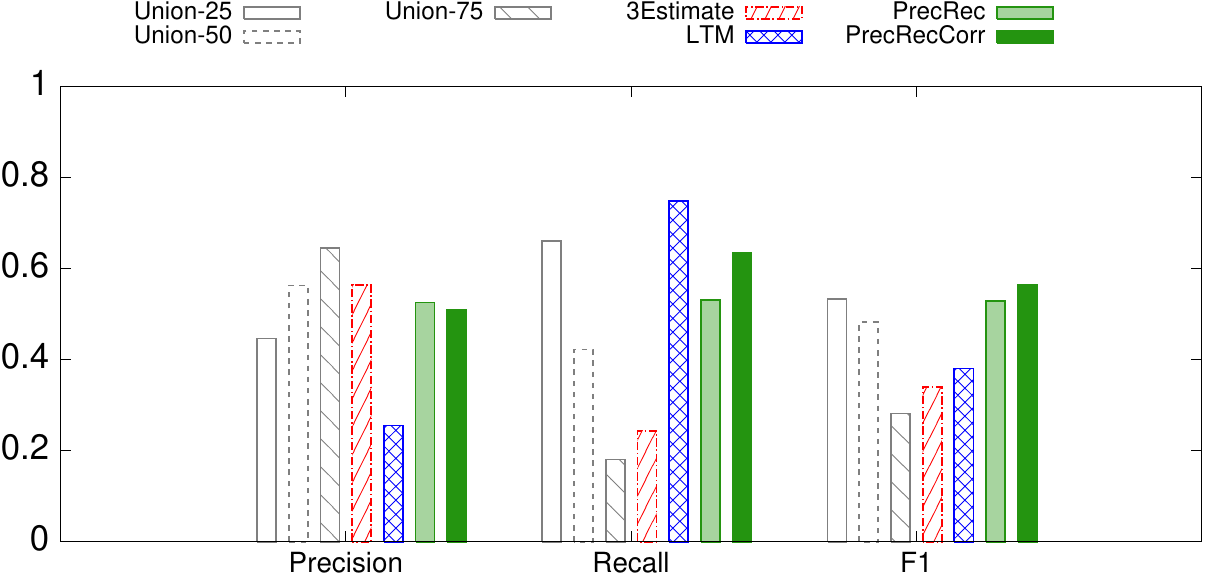}
\includegraphics[scale=0.7]{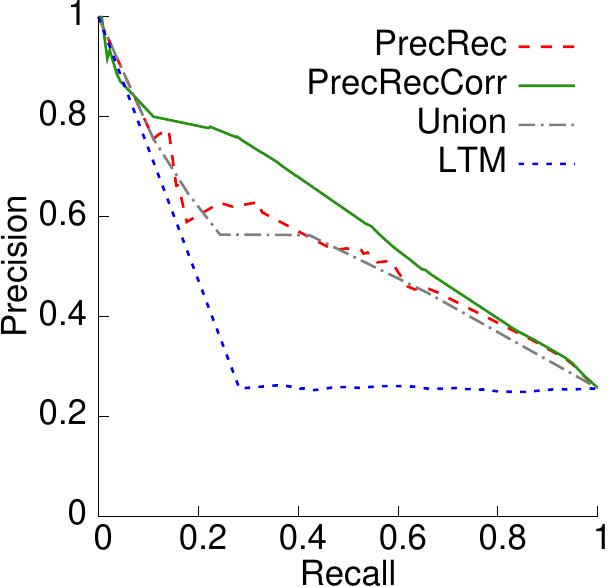}
\includegraphics[scale=0.7]{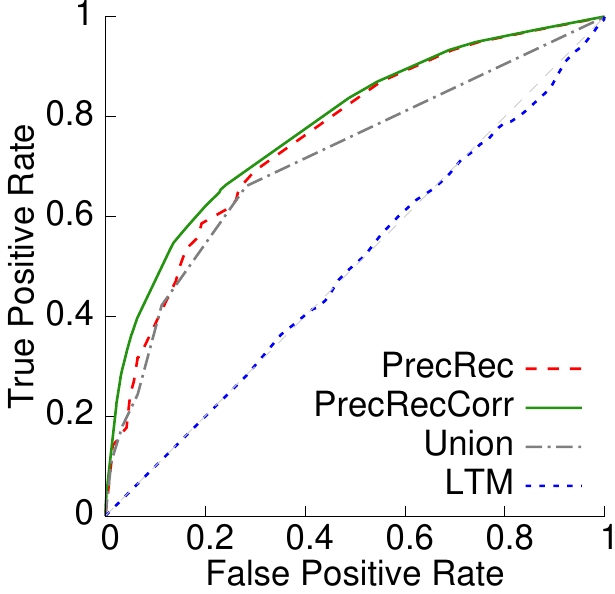}
\caption{Fusion results, and Precision-Recall and ROC curves for the \reverb data set.}\label{fig:reverb}
\end{subfigure}%

\begin{subfigure}[b]{\linewidth}
\centering
\includegraphics[scale=0.7]{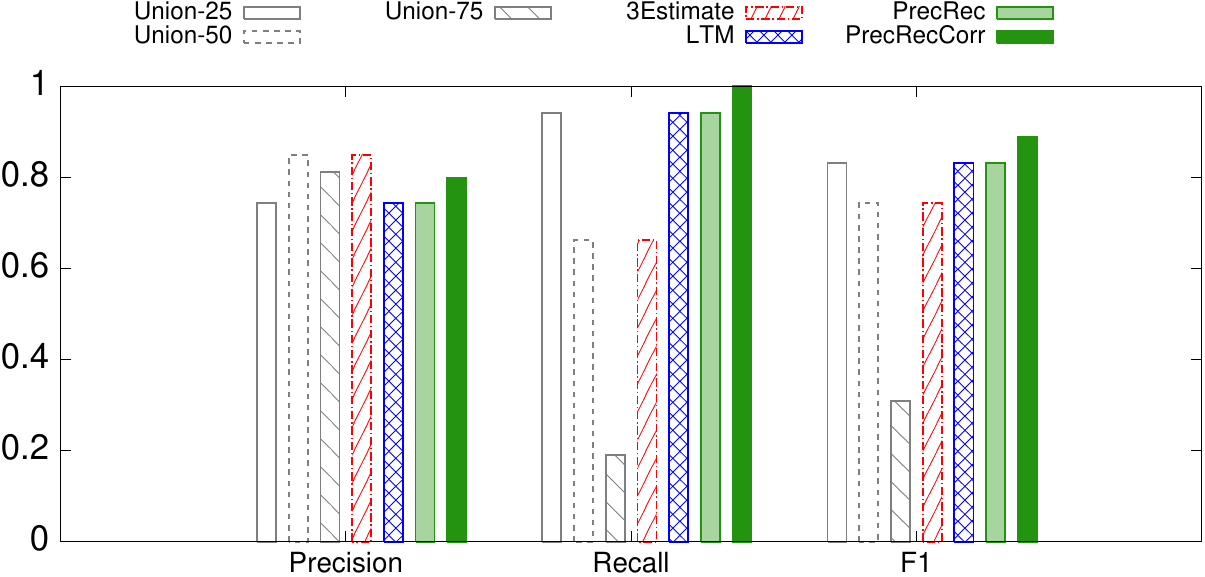}
\includegraphics[scale=0.7]{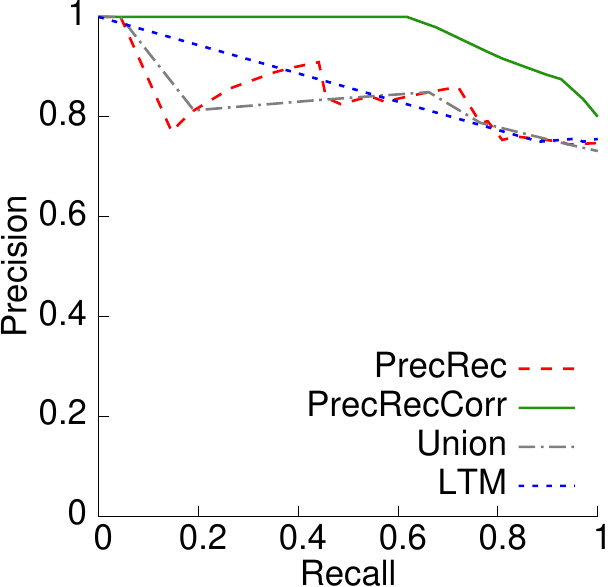}
\includegraphics[scale=0.7]{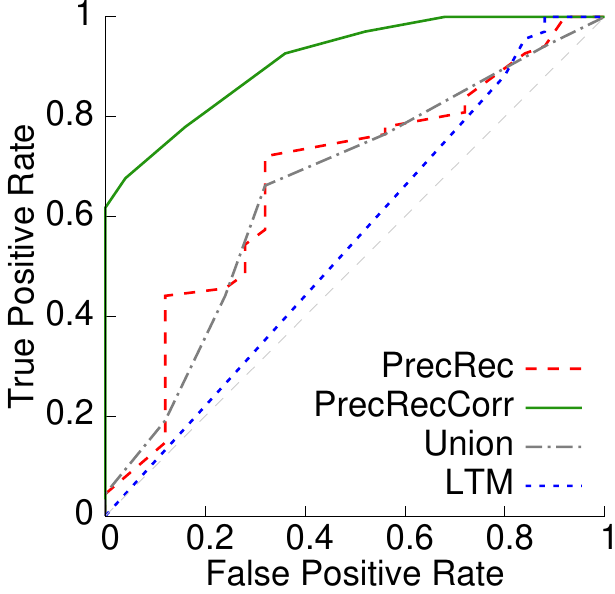}
\caption{Fusion results, and Precision-Recall and ROC curves for the \restaurant data set.}\label{fig:restaurant}
\end{subfigure}

\begin{subfigure}[b]{\linewidth}
\centering
\includegraphics[scale=0.7]{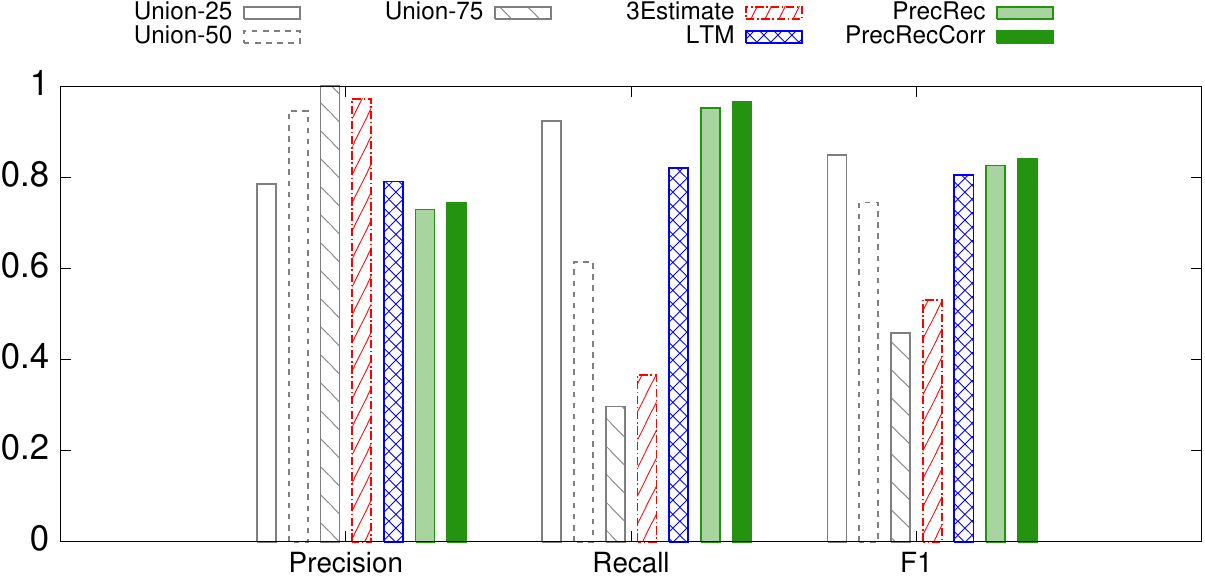}
\includegraphics[scale=0.7]{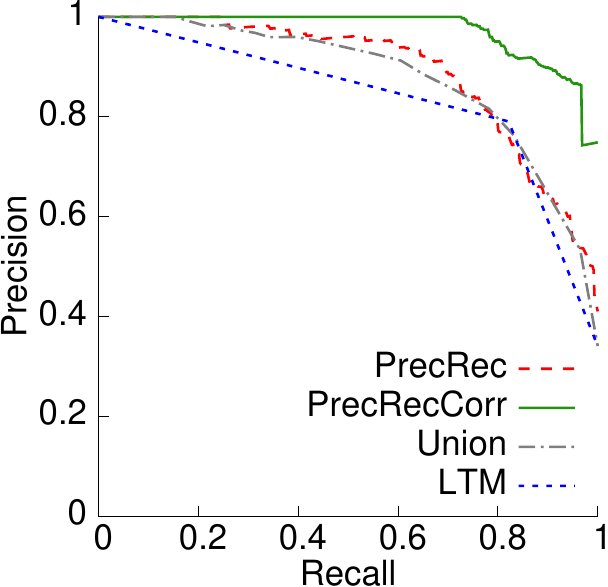}
\includegraphics[scale=0.7]{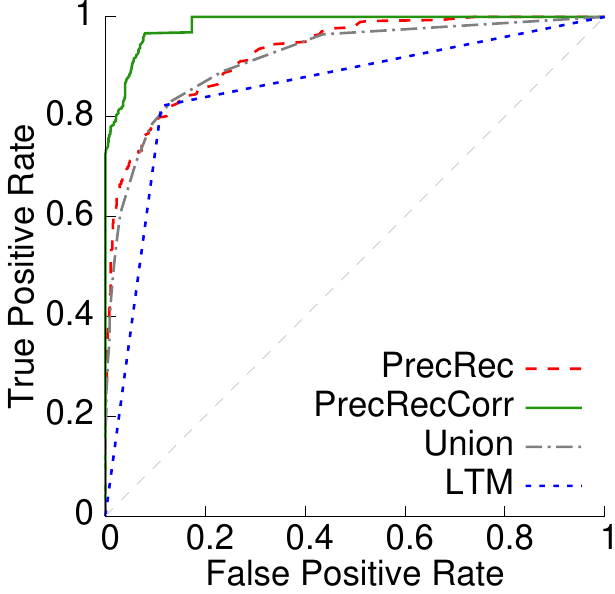}
\caption{Fusion results, and Precision-Recall and ROC curves for the \book data set.}\label{fig:book}
\end{subfigure}

\vspace{-1mm}
\caption{Our experiments show that \precrec and \preccorr result in better fusion results compared to other approaches.  In the \reverb dataset both \precrec and \preccorr showed significant improvement in the F-measure compared to the state-of the art (\estimate, and \ltm).  
In the \restaurant and \book datasets, \ltm and \union-25 are comparable to the results of \precrec, but the PR and ROC curves demonstrate that \preccorr provides significantly better truthfulness estimates for triples.}\label{fig:compare}
\vspace{-2mm}
\end{figure*}
}

\subsubsection*{Comparisons}

We compared our models with several state-of-the-art techniques that apply to the independent-triple and open-world semantics. 

\eat{
 \vspace{1mm}
\noindent
 \textbf{Majority:} Simple majority voting over the set of all sources. A triple is considered true if at least 50\% of the sources provide it.
}

 \vspace{1mm}

 \noindent
 \textbf{\union}-$K$: Considers a triple to be true if at least $K\%$
 of the sources provide it. Union-50 is equivalent to majority voting.
 \vspace{1mm}

\eat{
 \noindent
 \textbf{\cosine}~\cite{Ameli2010}: Compute source trustworthiness as the cosine similarity between provided data and truth-finding decisions, and iteratively compute trustworthiness of sources and truthfulness of triples. {\em Keep it???}
 \vspace{1mm}
}

 \noindent
\textbf{\estimate}~\cite{Ameli2010}: Iteratively computes trustworthiness of sources, trustworthiness of triples, and truthfulness of triples. This is the best model among the three proposed in~\cite{Ameli2010}, and we observed similar results from the other two models on our datasets.
 \vspace{1mm}

 \noindent
 \textbf{\ltm}~\cite{ltm2012}: Constructs a graphical model and uses Gibbs sampling to determine \newtext{source quality}\remove{recall and false positive rate of each source,} 
 and truthfulness of each triple. We used the default parameters suggested by~\cite{ltm2012}.
 \vspace{1mm}

\eat{
prior false positive count of a source = 10000
prior true negative count of a source = 100;
// these priors are used in beta distribution to generate false-positive-rate (FPR) value of a source

prior true positive count of a source = 50
prior false negative count of a source = 50 
// these priors are used in beta dist. to generate sensitivity of a source

prior true count of a fact = 10;
prior false count of a fact = 10;
// these priors are used in beta dist. to generate the prior truth prob. of a tuple. 
}

\eat{
\noindent
\textbf{\solomon}~\cite{solomon2009}: Detect copying between sources and discount votes from copiers. Note that it assumes conflicting-triple and closed-world semantics, so only applies to the \book\ dataset.
}

\noindent
\textbf{\precrec} (Section~\ref{sec:fusion_indep}): Computes truthfulness of each triple from the precision and recall of each source. We set $\alpha=0.5$ 
and computed source precision and recall according to the gold standard.
 
\vspace{1mm}
\noindent
\textbf{\preccorr} (Section~\ref{sec:fusion_corr}): Extends \precrec by considering correlation between sources. By default we report the results for the exact solution; however, as we show in Figure~\ref{fig:figs_levels}, we obtain similar results using level-3 elastic approximation. We computed joint precision and recall according to the gold standard. Note that \book is considerably larger than the other two datasets, which poses challenges for deriving the correlation parameters: (a) the number of correlation parameters is very large, and (b) there may not be enough support data to understand the correlation among the sources.  We overcome this issue using a simple clustering approach: we divide sources into clusters based on their pairwise correlations,
and assume that sources across clusters are independent. 

We used a C\# implementation of \ltm and we implemented the other models in Java.
For \reverb, \restaurant, and synthetic data, we ran experiments on a Macbook
Air with 4GB RAM, 1.7 GHz Intel Core i5 processor, and OSX Lion 10.7.5.
The \book experiments were run on a m1.large Amazon EC2 server instance~\cite{aws}.

\subsubsection*{Metrics}
We present results according to three metrics.

 \vspace{1mm}
\noindent
{\bf Precision/Recall/F1:} We measure the correctness of binary decisions with three metrics. \emph{Precision} measures among the returned true triples, 
how many are indeed true; \emph{recall} measures among the provided true triples, 
how many are returned; \emph{F-measure} computes their harmonic mean
(\ie, $F1={ 2 \cdot prec \cdot rec \over prec + rec}$). 

\vspace{1mm}
\noindent
{\bf PR-curve/ROC-curve:} We rank the provided triples in decreasing order of the computed 
truthfulness score (for {\sc Union}-$K$, we rank in decreasing order of the number of providers).
As we add the triples gradually, {\em PR-curve} plots the precision versus the recall after adding
each triple and {\em ROC-curve} plots the true positive rate versus the false positive rate.
In addition, we compute the area under the curve, called {\em AUC-PR} and {\em AUC-ROC} respectively.
These curves and measures allow us to examine whether the correctness probabilities we compute
are consistent with the reality. 

\vspace{1mm}
\noindent
{\bf Execution time:} We report execution time for each method.


\subsection{Real-World Data}
\label{sec:realData}

\eat{
\disable{
\begin{figure*}[t]
\begin{subfigure}[b]{.5\linewidth}
\centering
\includegraphics[scale=0.7]{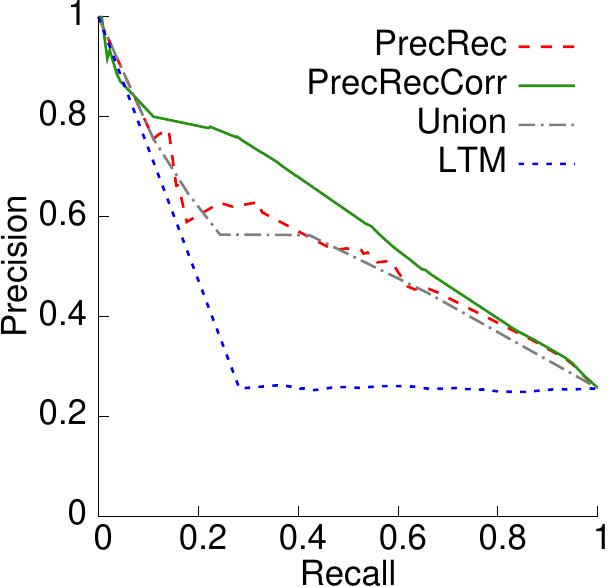}
\includegraphics[scale=0.7]{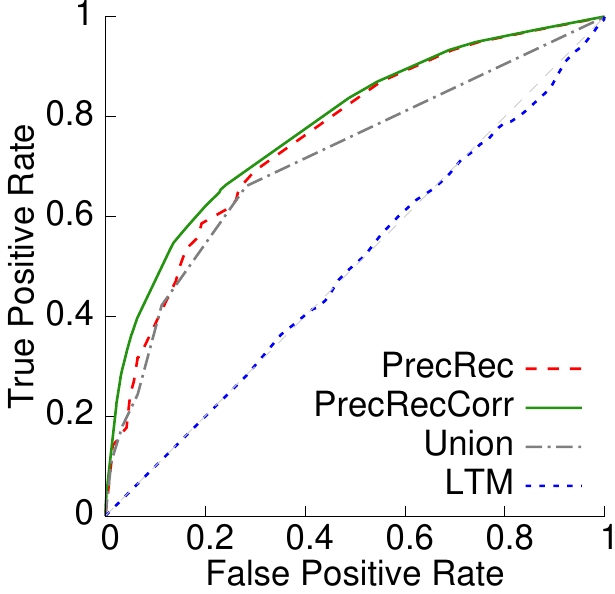}
\caption{Precision-Recall and ROC curves for the \reverb data set.}\label{fig:reverb-pr}
\end{subfigure}%
\hfill
\begin{subfigure}[b]{.5\linewidth}
\centering
\includegraphics[scale=0.7]{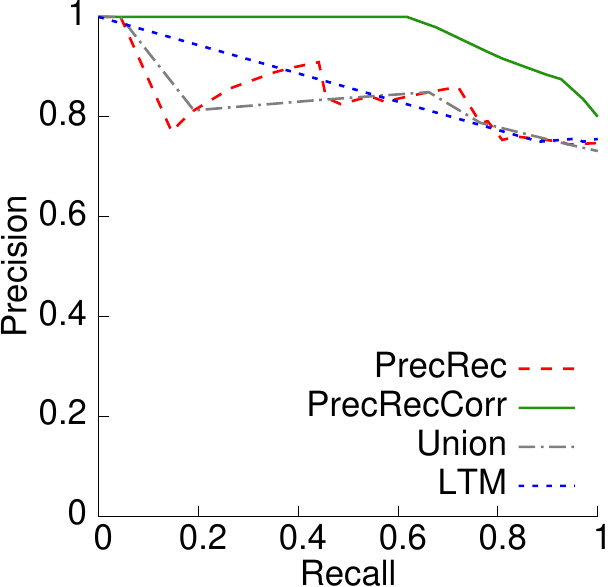}
\includegraphics[scale=0.7]{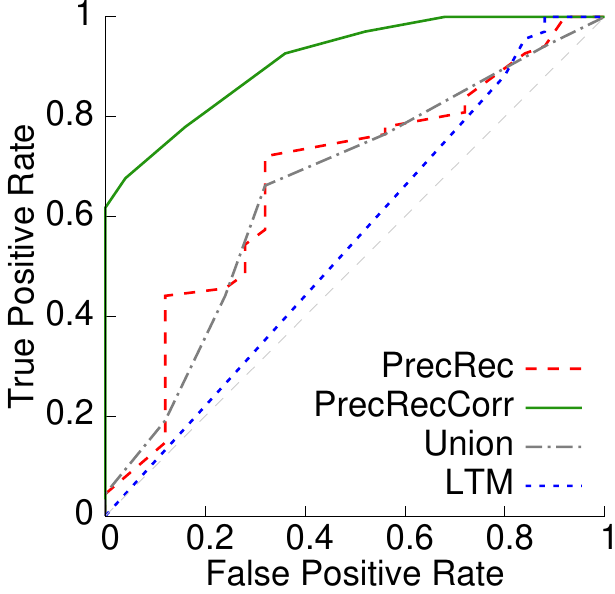}
\caption{Precision-Recall and ROC curves for the \restaurant data set.}\label{fig:restaurant-pr}
\end{subfigure}

\caption{The precision-recall and ROC curves demonstrate the gains of \preccorr over our basic approach that does not account for correlations (\precrec).  Both datasets demonstrate correlations.  However, \restaurant contains sources of much higher quality, which leads to better overall predictions.
}\label{fig:pr-curves}
\end{figure*}
}}

\begin{figure*}[th]
    \begin{subfigure}[b]{0.5\linewidth}
    \centering
        \includegraphics[scale=0.7]{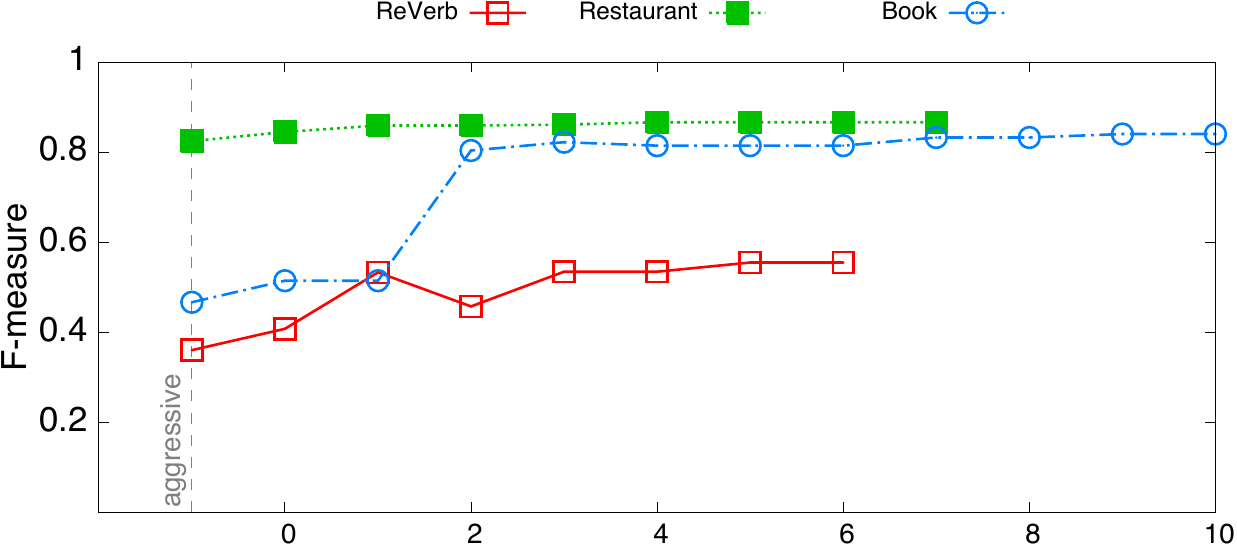}
    \caption{Elastic approximation levels}\label{fig:levelApprox}
    \end{subfigure}   
    \hfill
    \begin{subfigure}[b]{0.45\linewidth}
    \centering
    {\small
        \begin{tabular}{lrrr}
                \toprule
                \textbf{time(sec)}  & \reverb & \restaurant &\book\\
                \midrule
                \union-25       &    0.39      &   0.56       & 3.86\\
                \union-50       &   0.14       &   0.32      & 3.71\\
                \union-75       &   0.11   &   0.35      & 3.00\\
                \estimate       &   0.7   &   0.06     & 39\\
                \ltm (10 iter)  &   49   &   5.3        & 3791\\
                \precrec        &   2.6 &   0.3      & 35\\
                \preccorr       &   124       &   5.4      & 6786\\
                \preccorr-\textsc{lvl3}  &   79    &   2.25     & 2452\\
                \bottomrule
                \\
        \end{tabular}
    }
    \caption{Runtimes of algorithms (in seconds) for all datasets.}\label{fig:times}
    \end{subfigure}%
    \vspace{-2mm}
    \caption{As expected, our elastic approximation gradually approaches the result of \preccorr. 
    }\label{fig:figs_levels} 
\end{figure*}

We first compare the different models on the three real-world data sets. 
Figure~\ref{fig:compare} reports the precision, recall, and 
F-measure of each method on each dataset.
We also plot the PR-curve and ROC-curve of the methods on each data set.
Note that the curves for {\sc Union}-$K$ of different $K$ are the same
so we plot only one; also note that the results of \estimate\ 
are significantly worse than other methods, so we did not plot its curves to
avoid cluttering.


Overall, we observe that among different datasets, most of the methods obtain higher quality results
on \restaurant\ and \book, but lower quality on \reverb. This is not surprising
given that the data sources in \reverb have fairly low precision and recall and
they extract a lot of wrong triples.
\preccorr obtains the best results on all
datasets: comparing with \precrec, its F-measure is 5.2\% higher on average,
its AUC-PR is 10.3\% higher on average, and its AUC-ROC is 3.3\% higher on average.
We note that although the improvement on F-measure is not that large,
the improvement for AUC-PR and AUC-ROC is significant; this is because
with consideration of correlations between the sources, we often compute
a much higher probability for a true triple and a lower probability for a false triple, but this difference may be hindered when we apply the threshold
and make binary decisions.

\looseness -1
Among the methods that assume independence between sources,
\precrec obtains the best results: on average its F-measure is 14\% higher
than \ltm and 41\% higher than \estimate.
For \ltm, its F-measure is comparable to \precrec on \restaurant\ and \book,
but much lower on \reverb\ because of a very low precision. 
Its PR-curves and ROC-curves are not in a very good shape;
indeed, its AUC-PR is 24\% lower than \precrec and its AUC-ROC is 20.8\% lower
on average. We observed that the probabilities it outputs typically fall in extreme ranges; 
for example, for most of the triples that it considers as true on \restaurant, it computes a probability 
very close to 1.
\estimate\ obtains very low recall in all of the three datasets;
as a result, its F-measure is the lowest among all methods.

For \union-$K$, 
increasing $K$  increases the precision but drops the recall. 
\union-25 turns out to have the best F-measure, comparable
to \precrec on each data set, but lower than \preccorr.
However, its PR-curves and ROC-curves are in slightly worse shapes comparing with \precrec;
indeed, its AUC-PR and AUC-ROC is lower than that of \precrec by up to 4.5\%.
As we show later on synthetic data, \union-$K$ is sensitive on source quality;
for example, even \union-25 can obtain very low F-measure when the sources have
low precision or low recall. 


\looseness -1
Figure~\ref{fig:times} shows the execution time of the different models.
\union-$K$ is very efficient, while \estimate and \precrec\ are the next most
efficient, with runtimes up to one order of magnitude longer than
\union.
\newtext{
We terminated \ltm after 10 iterations; each iteration on average took 5.6
times longer than \precrec.
}
\preccorr is \remove{two}\newtext{one} order\remove{s} of magnitude slower than \remove{\union-$K$}\newtext{\precrec} on average;\remove{, but often shorter than \ltm.}
however, the level-3 elastic approximation obtained similar results 
but finished in only half of the time.
%
\newtext{
For our largest dataset (\book), level-3 approximated the exact solution in 40
minutes; we consider these runtimes reasonable, since this is an
offline cleaning process. Parallelization can significantly improve the
efficiency of \preccorr, as the terms at different levels and across different
clusters can be computed independently. With maximum parallelization \preccorr
terminates in 80 seconds, however a systematic study of these improvements is
outside the scope of this paper.
}

\disable{
\begin{figure*}[t]
\begin{subfigure}[b]{.3\linewidth}
\centering
\includegraphics[scale=0.65]{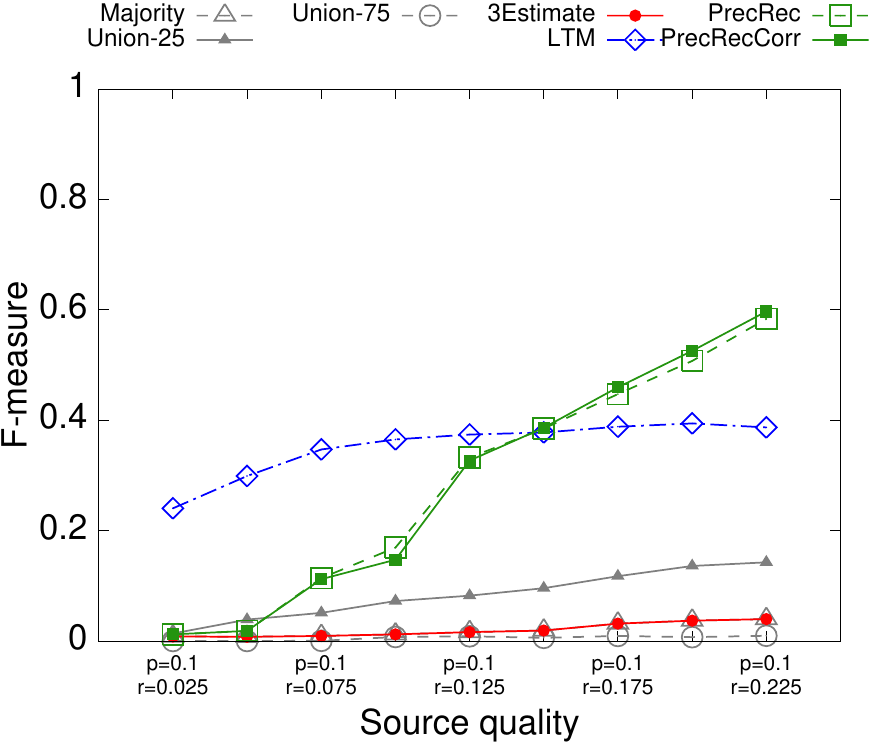}
\caption{Low precision sources, with low to fair recall, in a dataset of 25\% true triples.}\label{fig:badsources}
\end{subfigure}
\hfill
\begin{subfigure}[b]{.3\linewidth}
\centering
\includegraphics[scale=0.65]{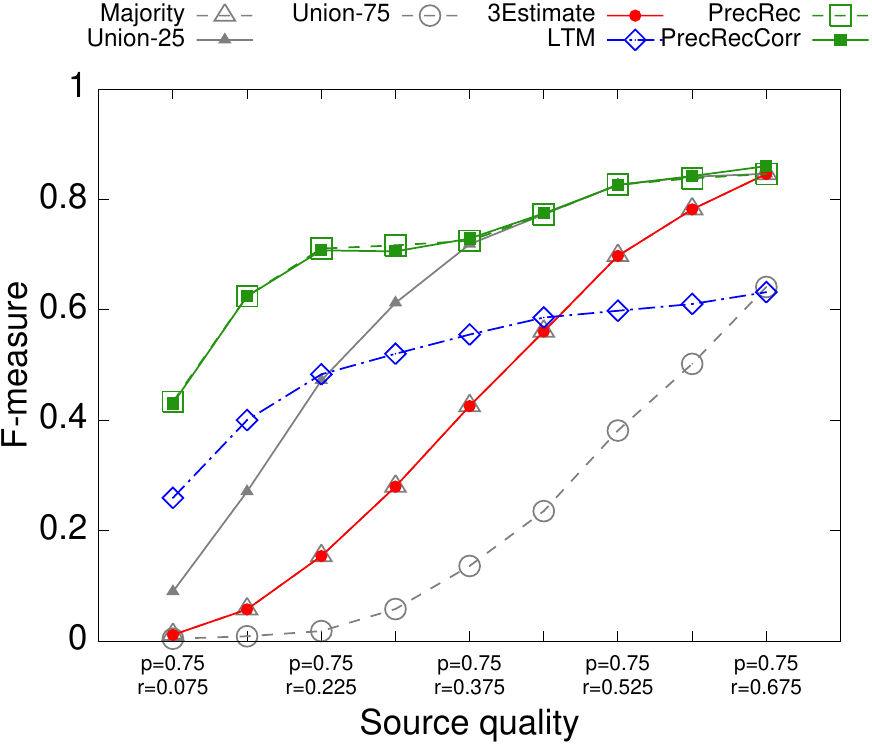}
\caption{High precision sources, with increasing recall, in a dataset of 50\% true triples.}\label{fig:goodVrecall}
\end{subfigure}
\hfill
\begin{subfigure}[b]{.3\linewidth}
\centering
\includegraphics[scale=0.65]{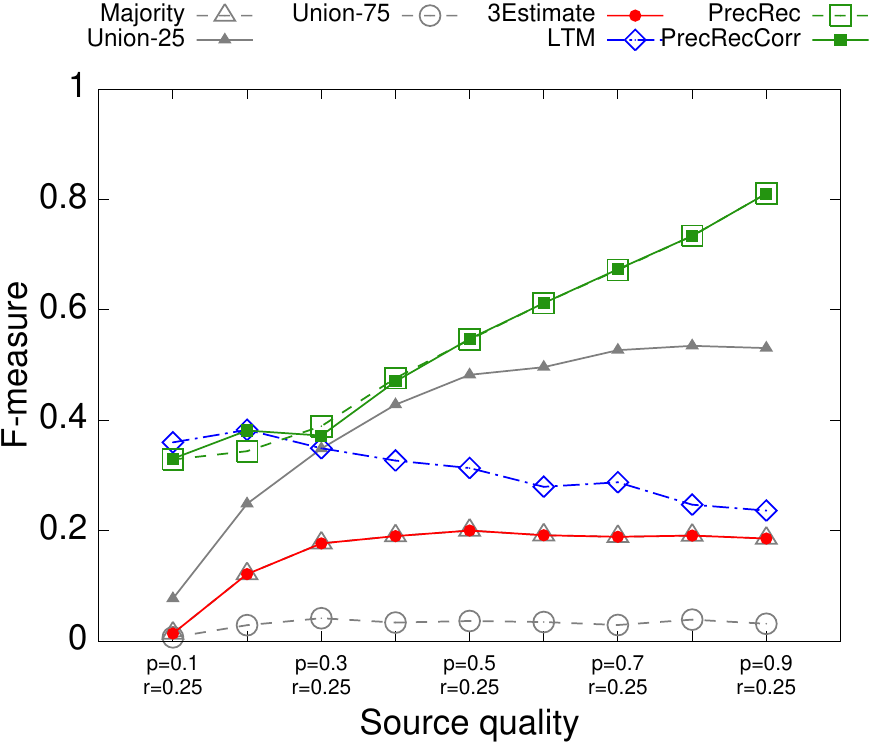}
\caption{Low recall sources, with increasing precision, in a dataset of 25\% true triples.}\label{fig:lowRecVprec}
\end{subfigure}
\vspace{-1mm}
\caption{Experimental results on synthetic data with independent sources. 
Our techniques are particularly effective with sources of low quality, and demonstrate significant gains in many configurations. 
\label{fig:indepSynthetic}}
\vspace{-2mm}
\end{figure*}
}

\smallskip
\noindent
\textbf{Elastic approximation:} 
Figure~\ref{fig:figs_levels} demonstrates the behavior of our aggressive
approximation and elastic approximation (Algorithm~\ref{alg:ptime_corr}) over
the three datasets. We observe that the aggressive estimate is much worse than
the exact solution on \reverb and \restaurant, while comparable on \book; it
is even worse than \precrec, which does not consider correlation. Each line in
the graph shows the progression of the approximation from the aggressive
estimate to the exact computation. At every level, the elastic approximation
refines the probability estimates of the earlier levels to gradually approach
\preccorr. Since the elastic approximation is heuristic in nature, there is no
guarantee that the method improves the estimate with every level (\eg, on
\reverb the elastic approximation performs worse at level 2 than level 1).
However, for all datasets, the elastic approximation comes close to the exact
result within a small number of levels.
We observe that on all three data sets, the result of level-3
approximation is already quite close to that of the exact solution, whereas
the execution time is much shorter.

\smallskip
\noindent
\textbf{Discovered correlations:} To better understand the improvement of \preccorr over \precrec,
we examine in more detail the discovered correlations between the sources. 

\reverb has 6 sources.
With respect to true triples, we detect strong correlation
on a group of 2 sources and on a group of 3 sources.
With respect to false triples, 2 pairs of sources are strongly correlated,
and one source is strongly anti-correlated with every other source.
Of the 7 \restaurant sources, 
we detect strong correlation on a group of 4 sources
and fairly strong anti-correlation on a pair of sources, with respect to true triples.
For false triples, there is strong correlation on a group of 6 sources. 
Finally, for \book, there are 333 sources that provide triples in the gold standard. 
Recall that we cluster the sources according to their correlation. 
In terms of true triples, we obtain three clusters of size 22, 3, and 2.
In terms of false triples, we obtain four clusters of size 22, 3, 2, and  2.
Interestingly, except two sources between which we find strong correlation both on true
triples and on false triples, the clusters for true triples and for false triples
contain very different sources. 

\looseness -1
These observations indicate that our model of correlation is much richer than what can be captured
by pure copying relationships, as in~\cite{solomon2009}.
For our datasets, \cite{solomon2009} applies only to \book\ dataset by considering the author list as a whole, but not the other datasets. In \book, this approach achieves high precision of 0.97 as it successfully detects copying and reduces the vote counts of false values. However, it has a low recall of 0.82, since it also discounts vote counts on true values and ignores other types of correlations. We leave an effective combination of that approach and ours for future work.

\subsection{Synthetic Data}
\label{sec:synthetic}

We generated  synthetic data to evaluate our algorithms under a large range of scenarios; in this section we present interesting cases that arise both in the case of independent sources, as well as in the case of correlations.

Our first set of experiments compares the different models on independent sources.
We generated 5 sources providing data on 1000 triples according to a pre-configured 
precision and recall; we averaged 10 repetitions and show the results in
Figure~\ref{fig:indepSynthetic}. 
Our results show that even without correlations, 
\precrec\ provides significant improvements over existing approaches, while \preccorr has similar performance.
Figure~\ref{fig:badsources} shows the performance of all the algorithms against a dataset of low quality sources. \ltm is quite robust to variations in source quality, and performs well in this challenging setting; however, it does not benefit much from increases in source quality, and \precrec quickly becomes better as recall increases over 0.15.
In Figures~\ref{fig:goodVrecall} and \ref{fig:lowRecVprec}, we vary recall and precision respectively, while keeping the other constant.  In both cases, our techniques perform remarkably well in comparison to the other algorithms. Note that \union-25 is very sensitive to source quality and performs badly with low-quality sources.

\begin{figure}[t]
    \centering
        \includegraphics[scale=0.7]{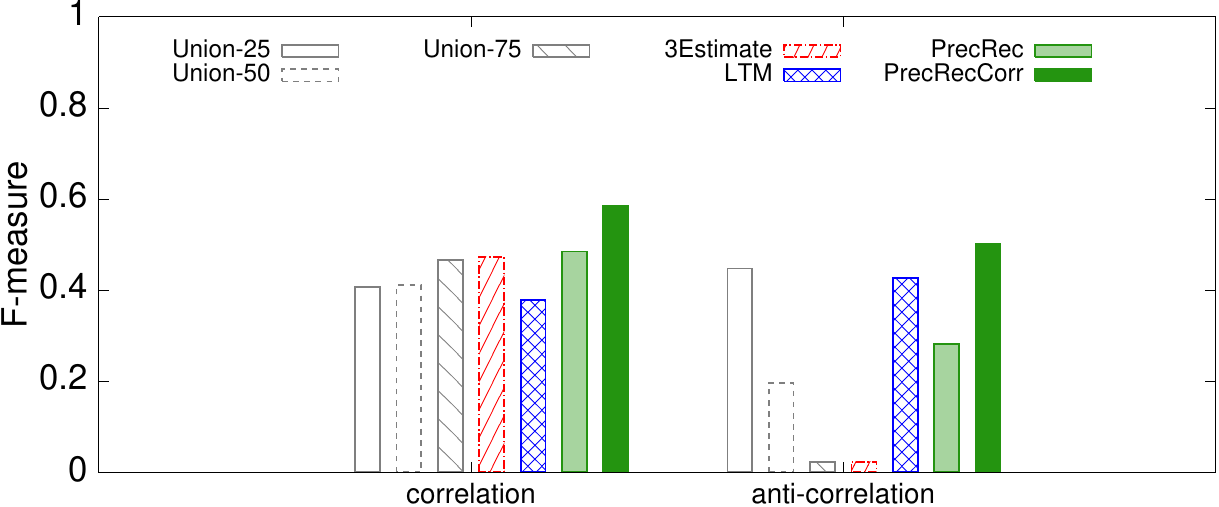}
    \caption{Experimental results on synthetic data with correlated sources. 
\preccorr obtains better results compared to all other approaches.}
    \label{fig:correlation}
    \vspace{-3mm}
\end{figure}

Our second set of experiments considers correlated sources. 
Figure~\ref{fig:correlation} demonstrates two cases: (1) a set of four sources are positively correlated on true triples, and (2) the sources are negatively correlated on false triples.  
In both cases, \preccorr\ demonstrates significantly better performance than all the other approaches.

\section{Related Work}
\label{sec:related}
There has been extensive work in the area of data fusion (\ie, resolving conflicts and
finding the truth); \cite{BDN07, DN09} surveyed early approaches and \cite{LDL+12}
compared recent approaches on Deep Web data. Among these approaches, 
\cite{solomon2009, Kleinberg98, PR10, PR11, PR13, YHY07, YT11} jointly infer truth 
and source quality, but they assume the conflicting-triple, closed-world semantics. 
{\sc Cosine} and {\sc 3Estimate}~\cite{Ameli2010} 
can be applied under the independent-triple, open-world semantics. Instead of using precision
and recall of sources, it considers a single quality metric--{\em accuracy} of a source;
we compared with them in our experiments (Section~\ref{sec:results}).
The model closest to ours is {\sc LTM}~\cite{ltm2012}; we have made detailed comparisons
in Section~\ref{sec:fusion_indep} and in experiments. All of these approaches assume independence
between sources. 

Correlation between sources are studied in two bodies of works. First, copy detection
has been surveyed in~\cite{DS11} for various types of data and studied 
in~\cite{BCM+10, Dong2010vldb, solomon2009, dong2009truth, LDOS11} for structured data.  
Our approach is different in three aspects. First, in addition to copying, we consider
broader scopes of correlations, including positive correlations not caused by copying
(\eg, extractors employing common extraction patterns), and also negative correlations.
Second, instead of just discounting votes from copiers, we may boost contributions
from providers correlated on true triples and reduce penalty from non-providers anti-correlated
on true triples. Third, we assume independent-triple and open-world semantics, 
opposite to their conflicting-triple, closed-world semantics. 
We have compared with this approach in our experiments. 

Second, there are other ways of measuring correlations. 
Qi et al.~\cite{QAH+13} constructed a graphical model that clusters dependent 
sources into groups and measures the quality of each group as a whole (instead of each individual source).
Kappa measure~\cite{kappa} measures correlation by taking into account 
the agreement by chance.
We measure correlations by the joint precision and recall for subsets of sources.
Our measures have much higher expressiveness in that
(1) they consider both positive and negative correlations;
(2) they distinguish correlation on true data and on false data;
and (3) they essentially consider correlation for every subset of sources.

\balance
\section{Contributions and Future Work} 
\label{sec:conclusion}
In this paper we presented a novel technique for fusing data that contains
correlations, which uses Bayesian analysis to derive the truthfulness of a
fact based on the quality of sources that provide it. We evaluated our
approach against other state-of-the-art techniques, and showed that our
algorithms achieve significant improvements in the fusion results. The power
of our approach lies in its generality: our algorithms do not need to have any
knowledge of possible correlations, and all required parameters can be
computed from a training set. As a result, \precrec and \preccorr perform well
even in low quality datasets that prove challenging for other techniques.

There are still several interesting challenges in this problem. Our model uses
independent-triple, open-world semantics, which allows our techniques to
consider multiple truth values for an entity (\eg, a person may have multiple
professions). However, this assumption may not always apply (\eg, a person
only has a single birth date). We consider modifications in our model to
account for such scenarios in future work. Another challenge is that source
quality may vary, based on the domain. For example, a source may have low
overall precision, but may be particularly accurate with respect to Pizzerias,
or restaurants in the Bay Area. In our model, we can consider domains
separately, but deriving the proper domain subdivisions automatically is not
straightforward.

\smallskip
\noindent
\textbf{Acknowledgements:}
We thank the authors of~\cite{ltm2012} for providing us with the implementation of \ltm.  This work was partially supported by NSF CCF-1349784 and a Google faculty research award.

{\small
\bibliographystyle{abbrv}
\bibliography{pdi}

\begin{thebibliography}{10}

\bibitem{aws}
{Amazon EC2 Instances}.
\newblock http://aws.amazon.com/ec2/instance-types/.

\bibitem{BDD+09}
L.~Berti-Equille, A.~D. Sarma, X.~Dong, A.~Marian, and D.~Srivastava.
\newblock Sailing the information ocean with awareness of currents: Discovery
  and application of source dependence.
\newblock In {\em CIDR}, 2009.

\bibitem{BCM+10}
L.~Blanco, V.~Crescenzi, P.~Merialdo, and P.~Papotti.
\newblock Probabilistic models to reconcile complex data from inaccurate data
  sources.
\newblock In {\em CAiSE}, 2010.

\bibitem{BDN07}
J.~Bleiholder, K.~Draba, and F.~Naumann.
\newblock {FuSem}-exploring different semantics of data fusion.
\newblock In {\em VLDB}, pages 1350--1353, 2007.

\bibitem{Dong2010vldb}
X.~L. Dong, L.~Berti-Equille, Y.~Hu, and D.~Srivastava.
\newblock Global detection of complex copying relationships between sources.
\newblock {\em PVLDB}, 3(1-2):1358--1369, Sept. 2010.

\bibitem{DBH+10a}
X.~L. Dong, L.~Berti-Equille, Y.~Hu, and D.~Srivastava.
\newblock Global detection of complex copying relationships between sources.
\newblock {\em PVLDB}, 2010.

\bibitem{solomon2009}
X.~L. Dong, L.~Berti-Equille, and D.~Srivastava.
\newblock Integrating conflicting data: The role of source dependence.
\newblock {\em PVLDB}, 2(1):550--561, 2009.

\bibitem{dong2009truth}
X.~L. Dong, L.~Berti-Equille, and D.~Srivastava.
\newblock Truth discovery and copying detection in a dynamic world.
\newblock {\em PVLDB}, 2(1), 2009.

\bibitem{DN09}
X.~L. Dong and F.~Naumann.
\newblock Data fusion--resolving data conflicts for integration.
\newblock {\em PVLDB}, 2009.

\bibitem{DS11}
X.~L. Dong and D.~Srivastava.
\newblock Large-scale copying detection.
\newblock In {\em Sigmod (Tutorial)}, 2011.

\bibitem{Fader11}
A.~Fader, S.~Soderland, and O.~Etzioni.
\newblock Identifying relations for open information extraction.
\newblock In {\em EMNLP}, 2011.

\bibitem{kappa}
J.~Fleiss.
\newblock {\em Statistical methods for rates and proportions}.
\newblock John Wiley and Sons, 1981.

\bibitem{Ameli2010}
A.~Galland, S.~Abiteboul, A.~Marian, and P.~Senellart.
\newblock Corroborating information from disagreeing views.
\newblock In {\em WSDM}, 2010.

\bibitem{Kleinberg98}
J.~M. Kleinberg.
\newblock Authoritative sources in a hyperlinked environment.
\newblock In {\em SODA}, 1998.

\bibitem{LDL+12}
X.~Li, X.~L. Dong, K.~B. Lyons, W.~Meng, and D.~Srivastava.
\newblock Truth finding on the deep web: Is the problem solved?
\newblock {\em PVLDB}, 6(2), 2013.

\bibitem{LDOS11}
X.~Liu, X.~L. Dong, B.~chin Ooi, and D.~Srivastava.
\newblock Online data fusion.
\newblock {\em PVLDB}, 4(12), 2011.

\bibitem{MW11}
A.~Marian and M.~Wu.
\newblock Corroborating information from web sources.
\newblock {\em IEEE Data Eng. Bull.}, 34(3):11--17, 2011.

\bibitem{mintz2009distant}
M.~Mintz, S.~Bills, R.~Snow, and D.~Jurafsky.
\newblock Distant supervision for relation extraction without labeled data.
\newblock In {\em ACL}, 2009.

\bibitem{PR10}
J.~Pasternack and D.~Roth.
\newblock Knowing what to believe (when you already know something).
\newblock In {\em COLING}, pages 877--885, 2010.

\bibitem{PR11}
J.~Pasternack and D.~Roth.
\newblock Making better informed trust decisions with generalized fact-finding.
\newblock In {\em IJCAI}, pages 2324--2329, 2011.

\bibitem{PR13}
J.~Pasternack and D.~Roth.
\newblock Latent credibility analysis.
\newblock In {\em WWW}, 2013.

\bibitem{QAH+13}
G.-J. Qi, C.~Aggarwal, J.~Han, and T.~Huang.
\newblock Mining collective intelligence in groups.
\newblock In {\em WWW}, 2013.

\bibitem{YHY07}
X.~Yin, J.~Han, and P.~S. Yu.
\newblock Truth discovery with multiple conflicting information providers on
  the web.
\newblock In {\em SIGKDD}, 2007.

\bibitem{YT11}
X.~Yin and W.~Tan.
\newblock Semi-supervised truth discovery.
\newblock In {\em WWW}, 2011.

\bibitem{ltm2012}
B.~Zhao, B.~I.~P. Rubinstein, J.~Gemmell, and J.~Han.
\newblock A {B}ayesian approach to discovering truth from conflicting sources
  for data integration.
\newblock {\em PVLDB}, 5(6):550--561, 2012.

\end{thebibliography}
}

\end{document}